\documentclass[fleqn,12pt]{wlscirep}

\usepackage[utf8]{inputenc}
\usepackage[version=4]{mhchem}
\usepackage[flushleft]{threeparttable}
\usepackage{booktabs}       
\usepackage{makecell}
\usepackage{siunitx}
\usepackage{setspace}
\usepackage{array}
\usepackage{silence}
\usepackage{bm}
\usepackage{wrapfig}
\usepackage{placeins}
\usepackage{fix-cm}

\newcommand{\PreserveBackslash}[1]{\let\temp=\\#1\let\\=\temp}
\newcolumntype{C}[1]{>{\PreserveBackslash\centering}p{#1}}
\newcolumntype{R}[1]{>{\PreserveBackslash\raggedleft}p{#1}}
\newcolumntype{L}[1]{>{\PreserveBackslash\raggedright}p{#1}}
\usepackage[T1]{fontenc}

\newcommand{\recheck}[1]{{\color{black}#1}}

\title{
A Graph Neural Network for the Era of Large Atomistic Models
}

\author[1,2,3]{Duo Zhang}
\author[1]{Anyang Peng}
\author[1,2]{Chun Cai}
\author[4]{Wentao Li}
\author[5,6]{Yuanchang Zhou}
\author[7,8,9]{Jinzhe Zeng}
\author[1,2,10]{Mingyu Guo}
\author[1,2,3]{Chengqian Zhang}
\author[11]{Bowen Li}
\author[12]{Hong Jiang}
\author[11,13,14]{Tong Zhu}
\author[5,6]{Weile Jia}

\author[1,2,\thanks{\href{mailto:linfeng.zhang.zlf@gmail.com}{linfeng.zhang.zlf@gmail.com}}]{Linfeng Zhang}
\author[15,16,\thanks{\href{mailto:wang_han@iapcm.ac.cn}{wang\_han@iapcm.ac.cn}}]{Han Wang}

\affil[1]{AI for Science Institute, Beijing 100080, P. R.~China}
\affil[2]{DP Technology, Beijing 100080, P. R.~China}
\affil[3]{Academy for Advanced Interdisciplinary Studies, Peking University, Beijing 100871, P. R.~China}
\affil[4]{Department of Chemical Engineering, Tsinghua University, Beijing 100084, P. R.~China.}
\affil[5]{State Key Lab of Processors, Institute of Computing Technology, Chinese Academy of Sciences, Beijing 100871, P.R.~China}
\affil[6]{University of Chinese Academy of Sciences, Beijing 100871, P.R.~China}
\affil[7]{School of Artificial Intelligence and Data Science, University of Science and Technology of China, Hefei 230026, P. R.~China}
\affil[8]{Suzhou Institute for Advanced Research, University of Science and Technology of China, Suzhou 215123, P. R.~China}
\affil[9]{Suzhou Big Data \& AI Research and Engineering Center, Suzhou 215123, P. R.~China}
\affil[10]{School of Chemistry, Sun Yat-sen University, Guangzhou, P. R.~China}
\affil[11]{Shanghai Engineering Research Center of Molecular Therapeutics \& New Drug Development, School of Chemistry and Molecular Engineering, East China Normal University, Shanghai 200062, P.R.~China}
\affil[12]{College of Chemistry and Molecular Engineering, Peking University
Beijing 100871, P. R.~China}
\affil[13]{NYU-ECNU Center for Computational Chemistry at NYU Shanghai, Shanghai 200062, P.R.~China}
\affil[14]{Institute for Advanced algorithms research, Shanghai, 201306, P.R.~China}
\affil[15]{National Key Laboratory of Computational Physics, Institute of Applied Physics and Computational Mathematics, Fenghao East Road 2, Beijing 100094, P.R.~China}
\affil[16]{HEDPS, CAPT, College of Engineering, Peking University, Beijing 100871, P.R.~China}

\begin{abstract}

Foundation models, or large atomistic models (LAMs), aim to universally represent the ground-state potential energy surface (PES) of atomistic systems as defined by density functional theory (DFT). 
The scaling law is pivotal in the development of large models, suggesting that their generalizability in downstream tasks consistently improves with increased model size, expanded training datasets, and larger computational budgets.
In this study, we present DPA3, a multi-layer graph neural network founded on line graph series (LiGS), designed explicitly for the era of LAMs.
We demonstrate that the generalization error of the DPA3 model adheres to the scaling law. 
The scalability in the number of model parameters is attained by stacking additional layers within DPA3.
Additionally, the model employs a dataset encoding mechanism that decouples the scaling of training data size from the model size within its multi-task training framework.
When trained as problem-oriented potential energy models, the DPA3 model exhibits superior accuracy in the majority of benchmark cases, encompassing systems with diverse features, including molecules, bulk materials, surface and cluster catalysts, two-dimensional materials, and battery materials.
When trained as a LAM on the OpenLAM-v1 dataset, the DPA-3.1-3M model exhibits lowest overall zero-shot generalization error across 12 downstream tasks spanning a diverse array of research domains.
This performance suggests superior accuracy as an out-of-the-box potential model, requiring minimal fine-tuning data for downstream scientific applications.

\end{abstract}
\begin{document}

\flushbottom
\maketitle
%
%
\thispagestyle{empty}


\section{Introduction}

The foundational model for atomistic systems is predicated on the Schrödinger equation~\cite{schrodinger1926quantisierung}, with the assumption that relativistic effects are negligible.
The ground state solution to the Schrödinger equation, applying the Born-Oppenheimer approximation~\cite{born1985quantentheorie}, defines a universal potential energy surface (PES), which plays a central role in the computational simulation of systems at the atomistic scale~\cite{frenkel2001understanding}.
In practice, Kohn-Sham density functional theory (DFT)~\cite{hohenberg1964inhomogeneous,kohn1965self} is frequently employed as a computationally feasible approximation to the ground state Schrödinger equation.
While DFT offers an attractive balance between accuracy and efficiency for most applications, the computational demand remains significant, as the complexity scales cubically with the electronic degrees of freedom.

In the past decade, machine learning interatomic potentials (MLIPs)~\cite{behler2007generalized,bartok2010gaussian,zhang2018deep,schutt2017schnet,batzner20223,gasteiger2021gemnet,batatia2022mace}  have emerged as an efficient surrogate for DFT calculations, significantly reducing computational costs to linear scaling while maintaining comparable accuracy~\cite{jia2020pushing}. 
MLIPs are typically trained to address specific scientific challenges. 
However, when investigating a new system, the model must be re-parameterized, necessitating a substantial amount of DFT calculations to label the training data.
This requirement has stimulated the development of universal models or large atomistic models (LAMs), which aim to universally represent the ground-state potential energy surface of the atomistic systems~\cite{chen2022universal,zhou2022uni,choudhary2023unified,deng2023chgnet,merchant2023scaling,batatia2023foundation,yang2024mattersim,rhodes2025orb,mazitov2025pet}. 
The feasibility of LAMs is grounded in the universality of the DFT solution. 
These models are expected to be used as out-of-the-box potential energy surfaces~\cite{batatia2023foundation} or fine-tuned with substantially less training data~\cite{shoghi2023molecules,zhang2024dpa2,wang2025pfd} for downstream tasks.


While some LAMs have been successfully applied, yielding substantial advances in research—such as the GNoME model, which discovered 381,000 new stable structures—a considerable disparity remains in the generalizability of state-of-the-art LAMs compared to MLIPs specifically trained for particular problems~\cite{peng2025lambench}.
To bridge this gap, enhancements in LAM architectures are expected to focus on improving generalizability across diverse research domains.
Additionally, the development of LAMs is anticipated to adhere to scaling laws~\cite{kaplan2020scaling,bi2024deepseek, su2024unraveling,yuan2025foundation}, which propose that the generalizability of these models can be improved by systematically expanding the dataset size, model parameters, and computational budgets.
Despite the pivotal role of scaling laws in model development, their exploration within LAMs remains limited. 
Notably, GNoME demonstrated a scaling law with respect to the number of training data, while Uni-Mol2 \cite{ji2024uni}, a pretrained model for molecular sciences, illustrated scaling laws concerning data, parameters, and computational budgets.
However, a recent study also reveals the non-trivial nature of scaling law, particularly for GNN-based architectures where oversmoothing effects pose significant challenges~\cite{li2025scaling}.
The specific model architecture through which a LAM may exhibit these scaling laws remains an open question.



According to the scaling laws, a key strategy for enhancing the generalizability of LAMs is to incorporate extensive training datasets that span a broad spectrum of research domains.
However, inconsistencies arise due to variations in exchange-correlation (XC) functionals, discretization basis sets, and software implementations in DFT calculations, rendering available training data incompatible for merging to train LAMs.
A potential solution is to expand existing datasets with compatible DFT calculation settings. 
For example, the Open Materials 2024 dataset \cite{barroso2024open} maintains consistent settings, specifically the PBE/PBE+U \cite{perdew1996generalized} functional and plane-wave basis, with the Materials Project \cite{jain2013materials}, while significantly increasing data volume.
However, this extension may not be able to satisfy the requirements for XC functionals and basis sets in other domains, such as small molecules, where hybrid functionals and atomic basis sets are commonly used~\cite{kovacs2023mace,anstine2024aimnet2,eastman2024nutmeg}.
Alternatively, multi-task training offers a method to learn common knowledge across a wide array of datasets, irrespective of their DFT settings~\cite{jacobson2023leveraging,zhang2024dpa2}. 
Nonetheless, this approach faces scalability challenges, as the number of fitting heads needed to learn dataset-specific knowledge increases with the number of datasets.
Furthermore, the inconsistent XC preferences across different domains drive the development of multi-fidelity models~\cite{kim_sevennet_mf_2024,chen2025one}.

The LAMs are expected to produce physically meaningful results in molecular simulations, necessitating adherence to all physical laws inherent to the universal PES.
Specifically, LAMs should be smooth and conservative, ensuring energy conservation in microcanonical ensemble molecular dynamics (MD) simulations and maintaining Boltzmann distributions in canonical and isothermal-isobaric ensemble MD simulations~\cite{tuckerman2010statistical}. 
LAMs must also be invariant under translational and rotational transformations, thereby conserving linear and angular momentum, respectively, in accordance with Noether's theorem.
Additionally, LAMs should respect the indistinguishability of atoms of the same chemical species, maintaining invariance under their permutations, which is fundamental to quantum statistics. 
Recent studies indicate that smooth, conservative models demonstrate a higher correlation between force field accuracy and property predictions, a critical factor in downstream applications~\cite{fu2025learning}.

In this study, we present the DPA3 model architecture, a message-passing neural network built upon line graph series (LiGS) specifically designed for the forthcoming era of LAMs.
The model is designed to exhibit scaling laws, whereby its generalizability improves consistently with increases in model size, the volume of training data, and computational budgets.
Furthermore, the DPA3 model incorporates dataset encoding to differentiate between training datasets, enabling multi-task training across diverse datasets irrespective of their DFT settings, with the advantage that the overhead does not scale with the number of datasets.
Last but not least, the DPA3 model is rigorously aligned with all physical laws associated with the universal PES.

The advantages of the DPA3 model architecture are first demonstrated through MLIP tasks that are well-established in the literature, where DPA3 consistently outperforms state-of-the-art GNN models in most instances.
Subsequently, the scaling law of DPA3 in LAM tasks is validated through training on the MPtrj dataset, which has been used to train LAMs such as CHGNet~\cite{deng2023chgnet} and MACE-MP-0~\cite{batatia2023foundation}.
Additionally, the DPA3 model, trained on the OpenLAM-v1 dataset~\cite{openlam-data-v1-web}, yields the DPA-3.1-3M model, which surpasses state-of-the-art LAMs in zero-shot force field predictions across 12 downstream tasks.
This superior performance underscores its potential in addressing real-world scientific challenges.
All of the features position DPA3 as an exceptionally suitable candidate for the era of LAMs.

\section{Results}

\subsection{DPA3: a graph neural network on line graph series}

\begin{figure}
    \centering
    \includegraphics[width=0.98\linewidth]{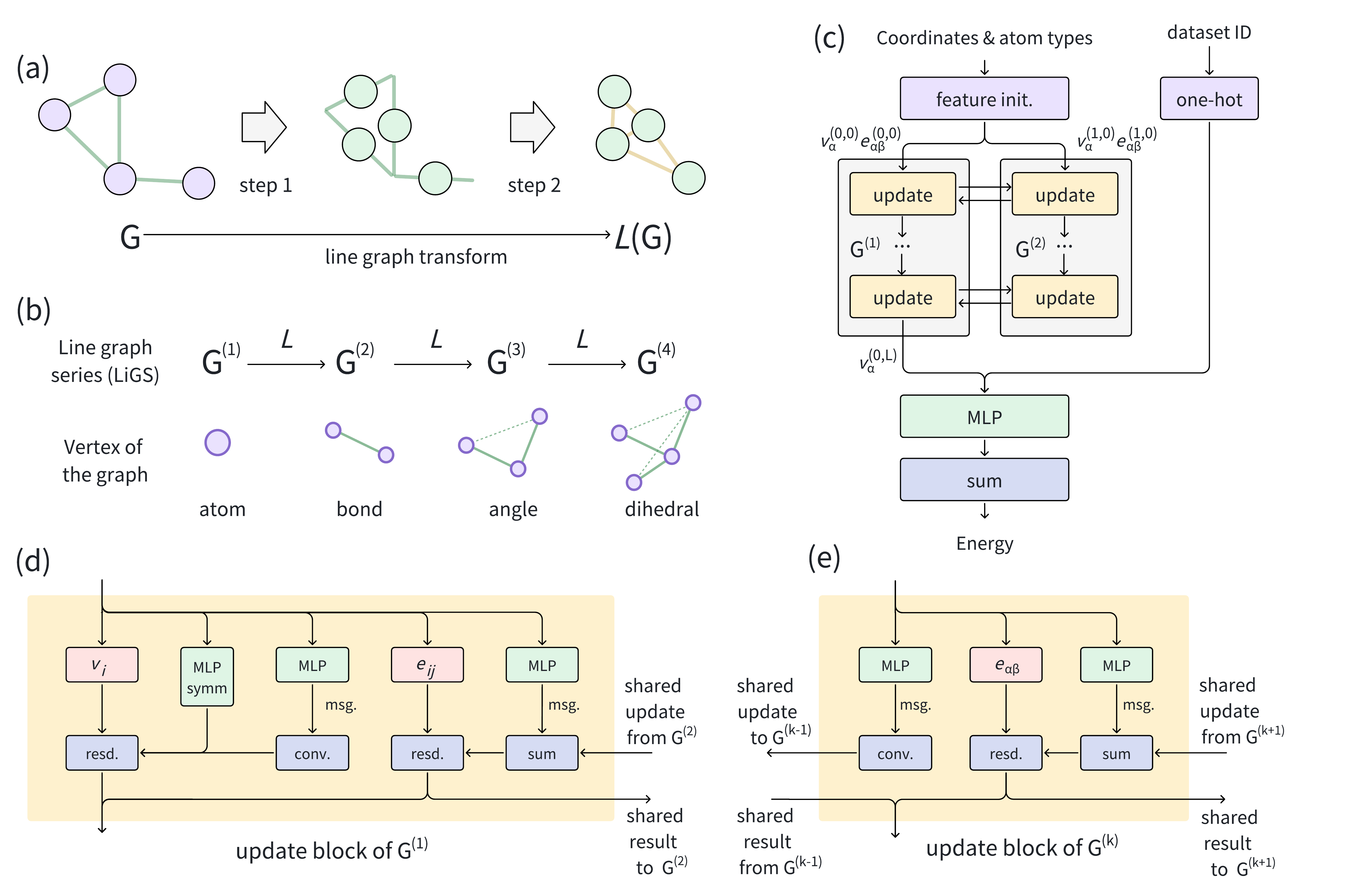}
    \caption{Schematic plot of the DPA3 model architecture. 
    (a) The line graph transform. 
    (b) The line graph series (LiGS). 
    (c) The model architecture of DPA3, a graph neural network on LiGS. 
    (d) The update block of graph $G^{(1)}$.
    (e) The update block of graph $G^{(k)}$, $k>1$.
    }
    \label{fig:DPA-3-arch}
\end{figure}

The DPA3 model is a graph neural network that operates on a series of graphs generated through the line graph transformation~\cite{whitney1992congruent,krausz1943demonstration,harary1960some}. 
Given a graph formed by vertices and edges, the line graph transform $\mathcal L$ constructs a new graph, denoted as \( \mathcal L(G) \). 
This process involves two steps, as illustrated in Fig.~\ref{fig:DPA-3-arch}(a): first, each edge in \( G \) becomes a vertex in \( \mathcal L(G) \); second, an edge is formed between two vertices in \( \mathcal L(G) \) if their corresponding edges in \( G \) share a common vertex.
It is important to note that applying the line graph transform to a graph results in another graph. Therefore, starting with an initial graph \( G^{(1)} \), one can recursively generate a series of graphs \(\{ G^{(1)}, G^{(2)}, \ldots, G^{(K)}\}\) using the line graph transform \( G^{(k)} = \mathcal L(G^{(k-1)}) \) for any \( 1 < k \leq K \). 
This sequence is referred to as the Line Graph Series (LiGS) generated from the graph \( G^{(1)} \), as illustrated in Fig.~\ref{fig:DPA-3-arch}(b).
In this series, we refer to the graph $ G^{(k)}  $ as having an order of $k$, while the maximal order $K$ also defines the order of the LiGS.    

In an atomistic system, a graph \( G^{(1)} \) can be defined by representing atoms as vertices and pairs of neighboring atoms as edges. 
The neighborhood of a given atom \( i \) consists of all other atoms within a user-defined cutoff radius, \( r^1_c \), from atom \( i \). 
A correspondence can be established between the vertices and edges in the LiGS and geometric entities within an atomistic system.
Specifically, the vertices in \( G^{(2)} \), \( G^{(3)} \), and \( G^{(4)} \) correspond to a bond defined by two neighboring atoms, an angle formed by three atoms with two of the bonds sharing a common atom, and a dihedral angle defined by four atoms with two of the angles sharing a common bond, respectively, as illustrated in Fig.~\ref{fig:DPA-3-arch}(b).

The DPA3 model is a multi-layer message-passing neural network defined on the LiGS, as depicted in Fig.~\ref{fig:DPA-3-arch}~(b).
In this model, the layer \( l \) vertex and edge features on the LiGS graph \( G^{(k)} \) are represented by \( v_\alpha^{(k,l)} \in \mathbb R^{d_v} \) and \( e_{\alpha\beta}^{(k,l)} \in \mathbb R^{d_e} \), respectively, where \( \alpha \) and \( \alpha\beta \) denote vertex and edge indices, $d_v$ and $d_e$ denote the dimensionality of the vertex and edge features, respectively. 
The feature updating process within the LiGS is conducted iteratively across all graphs using a recursive formula. 
At each layer, vertex features are refined through the convolution of messages transmitted via all edges connecting to the vertex, while edge features are updated based on a message formed from the edge feature itself combined with the features of the two terminal vertices. 
The updating mechanism employs a residual formulation to ensure model stability, even with the integration of multiple update layers.
Importantly, the vertex feature of graph \(G^{(k)}\) is identical to the edge feature of the preceding graph \(G^{(k-1)}\) within the LiGS framework.
This identity eliminates the necessity of preserving redundant vertex feature data for any graph \(G^{(k)}\) where \(k > 1\). 
Instead, vertex feature updates are transferred to adjacent graphs to revise the edge features of \(G^{(k-1)}\).
Subsequently, these updated edge features in graph \(G^{(k-1)}\) dictate the vertex features of graph \(G^{(k)}\), as illustrated in Fig.~\ref{fig:DPA-3-arch}(e).
For the initial graph \( G^{(1)} \), vertex features are iteratively refined by incorporating additional self-message and symmetrization transformation, see Fig.~\ref{fig:DPA-3-arch}~(d).
The detailed information on the feature update scheme is provided in Sec.~\ref{sec:dpa3-arch}.

The final vertex feature of the initial graph \(G^{(1)}\) is employed as a descriptor to represent the local environment of any atom within the system. 
Additionally, it is possible to aggregate vertex features from the graphs in the LiGS using suitable pooling methods. 
This approach enhances model performance in shallow architectures, whereas a decline in performance is observed with deeper models.
Given that model capacity is primarily augmented through deeper networks, we exclusively utilize the vertex feature from \(G^{(1)}\) to construct the descriptor.

In the case of multi-task training,
the descriptor is further augmented with a dataset encoding, typically represented as a one-hot vector corresponding to the training datasets. 
This enhanced feature is then input to a fitting MLP to predict the atomic contributions to the energy, which are subsequently combined to determine the total system energy. 
Forces and virials are derived by back-propagating the DPA3 predicted system energy with respect to atomic coordinates and cell tensors, respectively. 
Consequently, the DPA3 model is inherently conservative. 
The proof demonstrating that the DPA3 model is smooth and invariant to translational, rotational, and permutation symmetry operations is provided in Sec.~\ref{sec:dpa3-arch}.

\subsection{Related work}

\recheck{

\textbf{Graph neural network models on LiGS for atomistic modeling.}
Graph neural networks (GNNs) are widely used to predict materials properties~\cite{xie2018crystal}. 
Line graph transformations generate higher-order graphs that capture edge- and angle-associated features, thereby extending model capacity; representative examples include ALIGNN~\cite{choudhary2021atomistic}, DimeNet++~\cite{gasteiger2020fast}, CrysToGraph~\cite{wang2024crystograph}.
Ruff et.~al.~introduced the nested GNN architectures coGN and coNGN, which update the edge features of the initial graph $G^{(1)}$ using a nested GNN defined on $G^{(2)}$, the second graph in the LiGS.
ALIGNN-FF~\cite{choudhary2023unified} and CHGNet~\cite{deng2023chgnet} are GNN-based  potential-energy models built on the LiGS truncated at order $K=2$.
Relative to ALIGNN-FF and CHGNet, DPA3 provides three key advances: (1) A systematic extension of the architecture to higher LiGS orders. (2) New design elements—including the SiLUT activation and a residual update scheme—that together yield a superior performance and clear scaling laws with respect to model size, dataset size, and compute. 
Ablation studies quantifying the contribution of these components to the observed scaling behavior are provided in Sec.~\ref{sec:sm:ablation} of the Supplementary Materials.
(3) Dataset encoding, which enables parameter-efficient multitask training across datasets from diverse research domains.


\textbf{Equivariant GNNs.}
Equivariant graph neural networks~\cite{batzner20223,batatia2022mace,musaelian2022learning,liao2023equiformerv2,park2024scalable,fu2025learning} constitute a broad class of models for representing potential energy surfaces. 
These models can be interpreted as GNNs defined on the LiGS ${G^{(1)}}$, with their expressive power arising from the incorporation of equivariant atomic features that transform according to truncated irreducible representations of the symmetry groups O(3) or SO(3).
The key distinctions between equivariant GNNs and DPA3 are as follows: (1) equivariant GNNs operate on LiGS of order 1, whereas DPA3 is theoretically defined on LiGS of arbitrary order $K$; and (2) equivariant GNNs employ equivariant atomic features, while DPA3 leverages invariant features associated with the nodes and edges of each graph within the LiGS.
}

\subsection{Benchmarking} 
\label{sec:bench}

\begin{figure}
    \centering
    \includegraphics[width=0.9\linewidth]{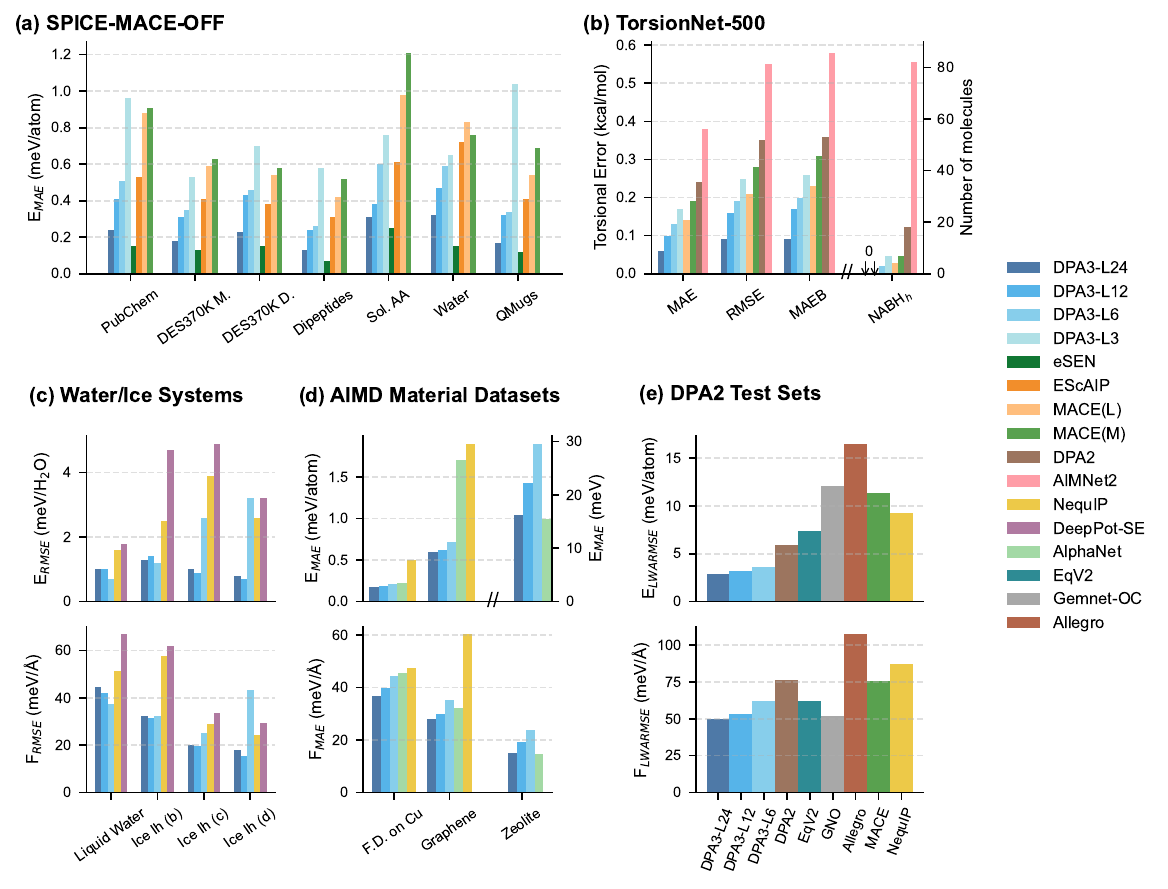}
    \caption{
    Comparative performance of DPA3 with other MLIPs across different benchmarks. 
    (a) Test energy MAE ($E_{MAE}$, meV/atom) evaluated on the SPICE-MACE-OFF dataset~\cite{kovacs2023mace}. 
    Abbreviations: DES370k M. (DES370k Monomers), DES370k D. (DES370k Dimers) and Sol. AA (Solvated Amino Acids). 
    (b) Torsional energy prediction errors quantified by MAE, RMSE, barrier height MAE (MAEB), and the number of accurately predicted barrier heights within 1 kcal/mol ($\mathrm{NABH}_h$) on the TorsionNet-500 dataset~\cite{rai2022torsionnet}, following evaluation methods described in Ref.~\cite{yang2024ab}. 
    (c) Test energy RMSE ($E_{RMSE}$, meV/\ce{H2O}) and force RMSE ($F_{RMSE}$, meV/\AA) for liquid water and three ice configurations (Ih (b), Ih (c), and Ih (d), sampled at different thermodynamic states)~\cite{zhang2018deep}.
    (d) Test energy MAE ($E_{MAE}$, meV/atom for the first two datasets and meV for the third) and force MAE ($F_{MAE}$, meV/Å) evaluated on three distinct material datasets from Ref.~\cite{yin2025alphanet}, namely, the Formate Decomposition on Cu (F.D. on Cu), the Defected Bilayer Graphene and the Zeolite dataset. 
    (e) Logarithmic Weighted Average RMSE of energy ($E_{LWARMSE}$, meV/atom) and force ($F_{LWARMSE}$, meV/\AA), as defined in Eq.~\eqref{eq:LWARMSE}, evaluated on the DPA2 test sets comprising 18 diverse cases detailed in Ref.~\cite{zhang2024dpa2}. 
    }
    \label{fig:result}
\end{figure}

The DPA3 model is evaluated using five test cases established in the literature, as depicted in Fig.~\ref{fig:result}.
In each case, the DPA3 models are trained as MLIPs, with a focus on achieving high accuracy in predicting energies, forces, and stress tensors. The results are then compared with those from state-of-the-art MLIP architectures.
In these benchmarks, we did not prioritize the accuracy of property calculations, primarily because the DPA3 model is designed to be smooth and conservative, thus convergence in force field error will inherently lead to convergence in property calculation error~\cite{fu2025learning}. 
In all instances, the maximum model size of the DPA3 is limited by the memory capacity of the GPU hardware accessible to the authors during the training period, specifically the Nvidia A800 GPU, which features 80GB of memory.
\recheck{For each benchmark dataset, we adopt the evaluation metric (MAE or RMSE) established in previous studies to enable a direct and equitable comparison with published baselines.
}

We first benchmark DPA3 on the SPICE-MACE-OFF~\cite{kovacs2023mace}, which was developed by Kov\'acs et.al.~for training a MLIP for organic small molecule tasks. 
The dataset comprises 1M configurations (95\% for training/validation and 5\% for test) with each labeled with the $\omega$B97M-D3(BJ)/def2-TZVPPD~\cite{najibi2018nonlocal, weigend2005balanced, rappoport2010property, grimme2011effect, grimme2010consistent} XC functional and basis set.
We have kept the division of the training, validation and test sets as the Ref.~\cite{kovacs2023mace}, and compare the DPA3 model with MACE-OFF23(M) (MACE(M) in short), MACE-OFF23(L) (MACE(L) in short), EScAIP~\cite{qu2024importance} and eSEN~\cite{fu2025learning} models. 
In this benchmark, we evaluate the DPA3 model with configurations of \recheck{3}, 6, 12, and 24 layers, denoted as \recheck{DPA3-L3}, DPA3-L6, DPA3-L12, and DPA3-L24, respectively. 
Additional hyper-parameters of the model are detailed in Section~\ref{sec:hyperparameters}.

To deliver a comprehensive evaluation of a model's performance on a test case consisting of multiple datasets, we introduce an averaged error metric calculated in logarithmic space: the Logarithmic Weighted Average Root Mean Square Error (LWARMSE) or Mean Absolute Error (LWAMAE),  defined as follows:
\begin{align}
\label{eq:LWARMSE}
\operatorname{LWARMSE \ or \ LWAMAE}(\{\mathbf{err}_i, w_{i}\}) = 
\operatorname{exp}\left(\frac{1}{\sum_{i}{w_i}}\sum_{i}{w_i \operatorname{log}(\mathbf{err}_i)}\right),
\end{align}
where $\mathbf{err}_i$ denotes the RMSE or MAE (of energy or force) calculated for the $i^{th}$ dataset within the test case,  and $w_i$ represents the corresponding weight assigned to each dataset. 
In the absence of explicitly specified weights, a default value of 1.0 is assigned to each subset. 
\recheck{This aggregate, computed in the logarithmic domain, mitigates the influence of outliers, preventing any single dataset from dominating the overall score and providing a more balanced assessment of performance across datasets.
}

As depicted in Fig.~\ref{fig:result}(a) and further detailed in the Supplementary Information Table~\ref{table:spice}, \recheck{our smallest model, DPA3-L3 attains the same energy LWAMAE as MACE(M) while using only 40\% of the parameters (0.9M vs 2.3M).
DPA3-L6, achieves approximately 34\% lower energy LWAMAE than MACE(L) with just 20\% of its parameters (1.3M vs 6.9M).
Remarkably, the larger DPA3-L24 configuration significantly surpasses MACE(L), achieving a reduction in energy LWAMAE by approximately 66\% while utilizing 30\% fewer parameters (4.9M vs. 6.9M).}
In comparison to EScAIP, 
DPA3-L24 exhibited superior performance across all energy MAE metrics and five out of seven force MAEs, utilizing only about 11\% of EScAIP's parameter count.
The eSEN model demonstrated the best performance among all models. 
The DPA3 family reveals a clear trend: as the model size increases, both energy and force errors decrease, suggesting that its performance may be further enhanced by employing deeper model configurations.

The accuracy of the SPICE-MACE-OFF trained DPA3 model in calculating the torsion profile is benchmarked using the TorsioNet-500~\cite{rai2022torsionnet} test case.
As illustrated in Fig.~\ref{fig:result}(b) and detailed in Table~\ref{table:torsion}, DPA3-L24 demonstrates superior performance across all metrics compared to models trained with SPICE-MACE-OFF, MACE(M/L), and other models such as AIMNet2~\cite{anstine2024aimnet2} and DPA2-drug~\cite{yang2024ab}.
Notably, it achieves a 60\% reduction in torsional barrier height MAE (MAEB) compared to MACE(L). 
Remarkably, both DPA3-L12 and DPA3-L24 exhibit zero NABH$_h$, indicating that the error in all predictions of torsional barrier height falls within the 1 kcal/mol threshold.

The performance of the DPA3 models in condensed-phase systems is preliminarily evaluated using a test case involving liquid water and ice Ih configurations. 
These configurations were derived from classical ab initio molecular dynamics (AIMD) and path-integral AIMD (PI-AIMD) simulations conducted under various thermodynamic conditions, utilizing the PBE0-TS XC functional~\cite{zhang2018deep}.
The liquid water configurations were sampled at ambient condition, while the ice Ih configurations were sampled at three thermodynamic states: state (b) 273~K and 1~bar, state (c) 330~K and 1~bar and state (d) 238~K and 2.13~kbar.
In accordance with the protocol established in previous research~\cite{batzner20223}, which utilized only 0.1\% of the data to demonstrate the data efficiency of the model, we adopted the same settings. 
This approach facilitates direct comparison among the DeepPot-SE~\cite{zhang2018end}, NequIP~\cite{batzner20223}, and DPA3 models.

As shown in Fig.~\ref{fig:result}(c) and detailed in Table~\ref{table:water}, DPA3-L12 achieves overall lower RMSE values compared to NequIP, with around 60\% reduction in energy LWARMSE and 30\% reduction in force LWARMSE. 
Notably, it even outperforms the original DeePMD model, which was trained on the complete dataset, in one out of four energy RMSEs and three out of four force RMSEs, underscoring DPA3’s strong fitting capability.
Interestingly, the scaling law is less apparent in this context: DPA3-L12 exhibits superior overall performance compared to DPA3-L24. This implies that the diversity of configurations within the test case is limited, causing larger models to be prone to overfitting.
Furthermore, our evaluation of DeepPot-SE—a refined version of the original DeePMD—under identical training conditions demonstrated comparable performance to NequIP, despite utilizing only 0.1\% of the data.
This finding challenges the validity of data efficiency assessments for this specific dataset~\cite{batzner20223}, emphasizing the necessity for more representative benchmarks to systematically evaluate model performance.


DPA3 was further evaluated using a test case proposed in Ref.~\cite{yin2025alphanet}, consisting of three datasets: formate decomposition on Cu, defected bilayer graphene, and zeolites. 
These datasets exemplify catalysis, two-dimensional materials, and porous materials, respectively.
As shown in Fig.~\ref{fig:result}(d) and Table~\ref{table:fcu}, the smallest model, DPA3-L6, consistently outperforms NequIP and demonstrates lower errors than AlphaNet, except for force prediction in the graphene dataset. 
Improved accuracy is achieved with larger models, DPA3-L12 and DPA3-L24, both of which consistently outperform NequIP and AlphaNet on the first two datasets.
For the zeolite dataset, a clear trend of reduced error rates is observed with larger DPA3 architectures.
The DPA3-L24 model demonstrates only slightly lower accuracy compared to AlphaNet. 

The DPA3 model is benchmarked on the DPA2 test sets~\cite{zhang2024dpa2}, which were originally introduced to evaluate the DPA2 model~\cite{zhang2024dpa2}. 
\recheck{DPA2 is a GNN constructed on the initial graph $G$ and employs a self-attention mechanism~\cite{vaswani2017attention} to model interactions among edge features. Consequently, its network architecture is fundamentally different from that of DPA3.}
These test sets comprise 18 datasets spanning diverse research domains, including alloy and battery materials, metal cluster catalysis, drug-like molecules, and linear alkane pyrolysis, all trained under an identical protocol. 
The LWARMSEs, calculated using the weights outlined in Table~\ref{table:single_task}, are depicted in Fig.~\ref{fig:result}(e). 
According to this metric, DPA3-L24 consistently demonstrates superior performance in both energy and force predictions, reinforcing its reliability across various applications.
Notably, EqV2 shows a lower force RMSE in certain systems, as detailed in Table~\ref{table:single_task}.
This may be attributed to its non-conservative force-prediction approach, which fits energy and force separately, at the expense of conservativeness.

The performance of DPA3 is also validated on well-established Matbench Discovery leaderboard \cite{riebesell2023matbench}, a benchmark that ranks MLIPs across various tasks simulating the high-throughput discovery of stable inorganic crystals. 
Specifically, compliant models trained on the MPtrj dataset~\cite{deng2023chgnet} are evaluated on the WBM test set~\cite{WBM} by performing structural optimization and predicting formation energies, subsequently converted to convex hull distances to assess thermodynamic stability.
Table~\ref{table:matbench} summarizes the comparative results between DPA3-L24 and other state-of-the-art compliant models accessed from the leaderboard on May 27, 2025. 
DPA3 achieves the second-best combined performance score (CPS), demonstrating an excellent balance between model complexity (parameter count) and predictive performance.
\recheck{Moreover, for DPA3, accuracy in energy prediction directly translates to accuracy in downstream property predictions. 
In Supplementary Information Section~\ref{sec:app:energy-property}, we demonstrate a strong linear correlation between the energy MAE on the WBM test set and the MAEs for four material properties—maximum phonon frequency, entropy, free energy, and heat capacity—in the MDR benchmark~\cite{loew2025universal}.
This correlation has been demonstrated to arise directly from enforcing strict conservativeness in the model architecture~\cite{fu2025learning}.
}

\begin{table}
  \scriptsize
  \caption{Results on the Matbench Discovery leaderboard, with all compliant models accessed before May 27, 2025.} 
  \label{table:matbench}
  \centering  
  \begin{threeparttable}
  \begin{spacing}{1.3}
  \begin{tabular}{L{2.4cm}R{0.7cm}R{0.7cm}R{0.7cm}R{0.7cm}R{0.7cm}R{0.7cm}R{0.7cm}R{0.7cm}R{0.8cm}R{0.8cm}R{0.8cm}R{0.8cm}}
    \toprule
    \textbf{Model} & \textbf{CPS}$\uparrow$ & \textbf{Acc}$\uparrow$ & \textbf{F1}$\uparrow$ & \textbf{DAF}$\uparrow$ & \textbf{Prec}$\uparrow$ & \textbf{MAE}$\downarrow$ & 
    \textbf{RMSE}$\downarrow$ & \textbf{R2}$\uparrow$ & \textbf{$\kappa$SRME}$\downarrow$ & \textbf{RMSD}$\downarrow$ & \textbf{Params}$\downarrow$ & \textbf{Targets} \\\hline
    DPA3-L24 & \underline{0.717} & 0.936 & 0.803 & 5.024 & 0.768 & 0.037 & \underline{0.080} & \underline{0.812} & 0.650 & 0.080 & 4.92M & EFSG \\\hline
    eSEN-30M-MP & \textbf{0.797} & \textbf{0.946} & \textbf{0.831} & \textbf{5.260} & \textbf{0.804} & \textbf{0.033} & \textbf{0.078} & \textbf{0.822} & \textbf{0.340} & \textbf{0.075} & 30.1M & EFSG \\
    SevenNet-l3i5 & 0.714 & 0.920 & 0.760 & 4.629 & 0.708 & 0.044 & 0.087 & 0.776 & 0.550 & 0.085 & 1.17M & EFSG \\
    MatRIS v0.5.0 MPtrj & 0.681 & 0.938 & 0.809 & \underline{5.049} & \underline{0.772} & 0.037 & 0.082 & 0.803 & 0.861 & \underline{0.077} & 5.83M & EFSGM \\
    GRACE-2L-MPtrj & 0.681 & 0.896 & 0.691 & 4.163 & 0.636 & 0.052 & 0.094 & 0.741 & \underline{0.525} & 0.090 & 15.3M & EFSG \\
    MACE-MP-0 & 0.644 & 0.878 & 0.669 & 3.777 & 0.577 & 0.057 & 0.101 & 0.697 & 0.647 & 0.091 & 4.69M & EFSG \\
    AlphaNet-MPTrj & 0.566 & 0.933 & 0.799 & 4.863 & 0.743 & 0.041 & 0.091 & 0.745 & 1.310 & 0.107 & 16.2M & EFSG \\
    eqV2 S DeNS & 0.522 & \underline{0.941} & \underline{0.815} & 5.042 & 0.771 & \underline{0.036} & 	0.085 & 0.788 & 1.676 & 0.076 & 31.2M & EFSD \\
    ORB v2 MPtrj & 0.470 & 0.922 & 0.765 & 4.702 & 0.719 & 0.045 & 0.091 & 0.756 & 1.725 & 0.101 & 25.2M & EFSD \\
    M3GNet & 0.428 & 0.813 & 0.569 & 2.882 & 0.441 & 0.075 & 0.118 & 0.585 & 1.412 & 0.112 & 228k & EFSG \\
    CHGNet & 0.400 & 0.851 & 0.613 & 3.361 & 0.514 & 0.063 & 0.103 & 0.689 & 1.717 & 0.095 & 413k & EFSGM \\
    \toprule
  \end{tabular}
  \end{spacing}
  \end{threeparttable}
\end{table}

\begin{figure}
    \centering
    \includegraphics[width=0.6\linewidth]{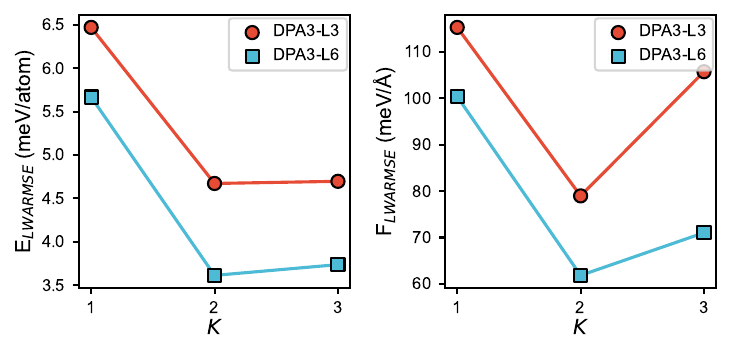}
    \caption{LWARMSEs in energy and force predictions evaluated on the DPA2 test sets. 
    The DPA3 models with three layers (DPA3-L3) and six layers (DPA3-L6) were examined at varying LiGS orders \(K\). 
}
    \label{fig:order}
\end{figure}

The performance of the DPA3 architecture, constructed with varying orders of the LiGS, is evaluated using the DPA2 test sets. 
Specifically, we assessed the LWARMSEs of both energy and forces for DPA3-L3 and DPA3-L6 under different LiGS orders \(K\), as illustrated in Fig.~\ref{fig:order}. 
The model's accuracy improves significantly when increasing the order from 1 to 2; however, this trend does not persist as the order increases to 3. 
In fact, a decrease in accuracy is observed when increasing the order from 2 to 3, particularly in the accuracy of force predictions. 
In theory, the order-2 DPA3 possesses a higher capacity than the order-3, which suggests the observed decline may be attributed to two factors: 
First, the edge features in the order-2 graph, geometrically represented as angular features, sufficiently capture the local geometric context for most systems. 
Second, the inclusion of edge features in the order-3 graph, specifically dihedral terms, potentially complicates the training process.
Based on these findings, we adopt order \(K=2\) as the default configuration, as it provides the optimal performance.

\recheck{A systematic assessment of inference performance was conducted, with detailed protocols provided in Section~\ref{sec:sm:effi}.
Performance is quantified as the average time per atom required to compute the energy, forces, and pressure tensor of a given configuration.
Because computational efficiency depends strongly on system size, we employed an extensible water system at equilibrium density as a representative example and evaluated DPA3 (L3, L6, L12, and L24) and MACE (M and L variants) across configurations of varying sizes (Fig.~\ref{fig:sm:effi}).
The runtimes of MACE and DPA3 converge beyond roughly 400 and 1000 atoms, respectively.
DPA3-L3 is the fastest among all models—about 2$\times$ faster than MACE(M)—while also achieving slightly better accuracy than MACE(M) in Table.~\ref{table:spice}.
DPA3-L6 attains a similar speed to MACE(M) (yet delivers 34\% better accuracy than MACE(L) ), and DPA3-L12 is faster than MACE(L).
Results for optimized MACE implementations with customized CUDA kernels~\cite{cuda_mace,kovacs2023mace} are also included.
For relatively small systems (fewer than $\sim$300 atoms for MACE(M) and $\sim$100 atoms for MACE(L)), the optimized MACE versions show no clear advantage, but their efficiency surpasses the standard implementations as system size increases.
At convergence, optimized MACE (M, L) outperforms both MACE(M) and DPA3-L6, and the optimized MACE(M) achieves a speed matching DPA3-L3.
}

\subsection{Scaling Law}
\label{sec:scaling-law}

\begin{figure}
    \centering
    \includegraphics[width=1.0\textwidth]{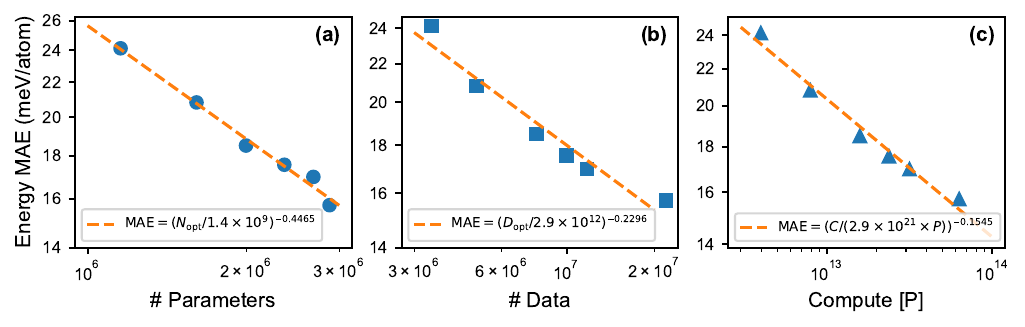}
    \caption{\recheck{Scaling law of the DPA3 models. 
    All evaluations are conducted by measuring validation energy MAEs using models trained on the OMat24 dataset. 
    DPA3 exhibits smooth performance improvement with jointly scaling of (a) model parameters ($N$), (b) training data ($D$), and (c) compute budget ($C$).}}
    \label{fig:scale}
\end{figure}

\recheck{The scaling behavior of DPA3 is examined on the OMat24 dataset using an approach analogous to the IsoFLOP profiles introduced in Ref.~\cite{hoffmann2022training} and subsequently applied to study the scaling laws of UMA models~\cite{wood2025umafamilyuniversalmodels}.
Training experiments were conducted with varying model sizes under six fixed IsoFLOP settings.
The total computational cost is estimated as $C = P \times N \times D$, where $C$ denotes the compute budget (in FLOPs), $N$ the number of model parameters, and $D$ the size of the training dataset (in number of configurations).
Here, $P$ is an unknown prefactor whose exact value is unimportant.
All experiments were performed in the single-epoch regime, such that $D$ is proportional to the number of training steps.
For each compute budget, the validation MAE was plotted as a function of parameter count $N$ and fitted with a parabola.
The minimum  MAE and the corresponding optimal parameter count, $N_{\textrm{opt}}(C)$, represent the best achievable performance under the given compute budget.
The validation MAE is further analyzed as a function of $N_{\textrm{opt}}$, the corresponding dataset size $D_{\textrm{opt}} = C / (P N_{\textrm{opt}})$, and the compute budget $C$, as shown in Fig.~\ref{fig:scale}(a–c).
The results clearly demonstrate that the validation energy MAE decreases with increasing model size, dataset size, and compute budget, following power-law scaling consistent with trends observed in large language models~\cite{kaplan2020scaling, hoffmann2022training}, albeit with different exponents.}

%

\subsection{A large atomistic model: DPA-3.1-3M}
\label{sec:LAM}

\begin{figure}
    \centering
    \includegraphics[width=0.9\textwidth]{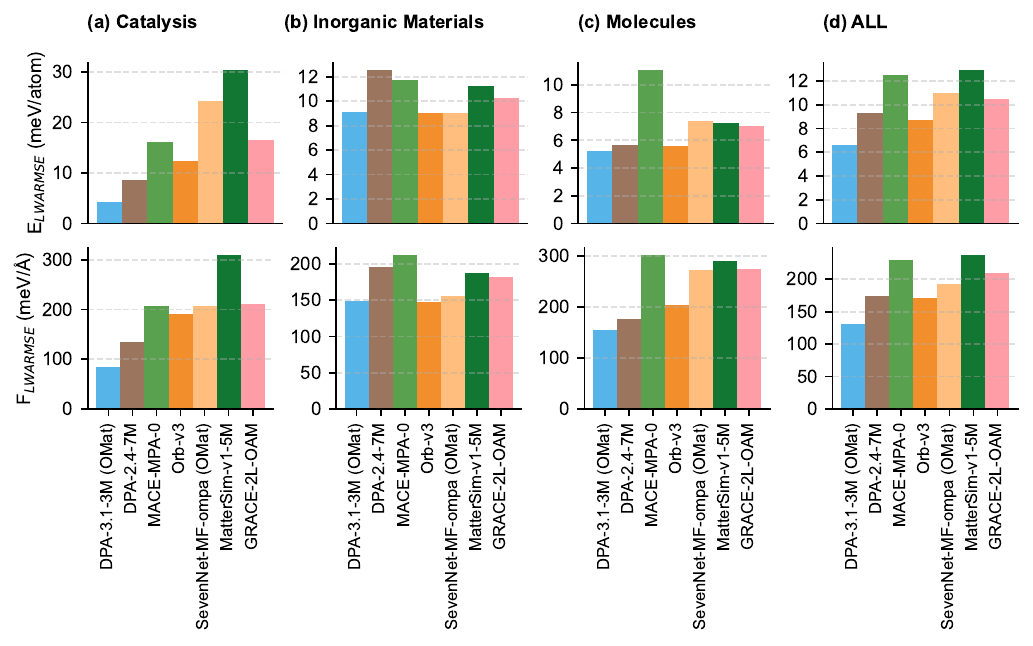}
    \caption{
    Zero-shot generalizability evaluation of force field prediction tasks across three distinct domains. 
    (a-c) Logarithmic Weighted Average Root Mean Squared Errors (LWARMSE) of energy ($E_{\text{LWARMSE}}$, meV/atom) and force ($F_{\text{LWARMSE}}$, meV/Å), defined by Eq.~\eqref{eq:LWARMSE}, computed separately within each domain. 
    (d) LWARMSE evaluated on the combined set of all 12 datasets across the three domains.
    \recheck{
    Training data summary:
    DPA-3.1-3M and DPA-2.4-7M~\cite{zhang2024dpa2,peng2025lambench} are multitask-pretrained on OpenLAM-v1~\cite{openlam-data-v1-web,peng2025lambench}.
    SevenNet-MF-ompa~\cite{kim2024data} is multitask-pretrained on OMat24~\cite{barroso2024open}, MPtrj~\cite{deng2023chgnet}, and sAlex~\cite{schmidt2023machine}.
    MACE-MPA-0~\cite{batatia2023foundation}, Orb-v3~\cite{rhodes2025orbv3}, and GRACE-2L-OAM~\cite{bochkarev2024graph} are pretrained on OMat24 and subsequently fine-tuned on MPtrj and sAlex.
    MatterSim-v1-5M~\cite{yang2024mattersim} is trained on a proprietary materials dataset that is not publicly available.
    }
    }
    \label{fig:fftask}
\end{figure}

\recheck{In this work, we propose a large atomistic model (LAM), DPA-3.1-3M, which is trained on the OpenLAM-v1 dataset collection. 
This collection consists of 31 datasets with random train/validation splits and includes large-scale open-source datasets such as OMat24~\cite{barroso2024open}, OC20~\cite{chanussot2021open}, and SPICE2~\cite{eastman2024nutmeg}, as well as smaller, domain-specific datasets (e.g., metals, alloys, and organic reactions), as documented in the data card~\cite{openlam-data-v1-web}.
To jointly train datasets with varying sizes and DFT settings, we adopted a multitask pretraining scheme~\cite{shoghi2023molecules, zhang2024dpa2}, utilizing per-dataset sampling weights that were heuristically selected based on dataset importance and sparsity. A detailed description of each dataset and the corresponding weights is available in Refs.~\cite{openlam-data-v1-web, peng2025lambench}.
The DPA-3.1-3M model architecture employs a LiGS order truncated at $K=2$ and consists of $L=16$ layers.
The model was trained for 4 million steps on 128 GPUs, with detailed hyperparameters provided in Table~\ref{table:hyperparam}. 
Under this setup, the resulting DPA-3.1-3M model contains approximately 3.26 million parameters.
}

The generalizability of the DPA-3.1-3M model was assessed by evaluating its zero-shot prediction accuracy on force field tasks across 12 downstream datasets spanning three distinct scientific domains: \textit{Catalysis}, \textit{Inorganic Materials}, and \textit{Molecules}, as detailed in Tab.~\ref{tab:ffperformance}.
These datasets were proposed in the literature as training resources for MLIPs designed to address various scientific challenges, and for efficiency, we randomly downsampled each to a maximum of 1,000 data frames.
Importantly, they were independently constructed from the training datasets within the OpenLAM-v1 collection.
In this context, zero-shot prediction refers to the direct evaluation of the DPA-3.1-3M model's prediction error on the downstream datasets without any additional training.
It is important to note that energy labels computed via DFT can differ by an arbitrary constant due to variations in pseudopotential selection and software implementations.
To address this, for energy prediction, each model in this test is employed to predict the energy difference between the labels and a dummy model that linearly correlates the DFT energy with chemical composition. 
In contrast, predictions for forces and virials are directly obtained from the models without such adjustments.

Figure~\ref{fig:fftask} and Table~\ref{tab:ffperformance} present a comparative analysis of the zero-shot generalizability of the DPA-3.1-3M model against several state-of-the-art LAMs released prior to May 1, 2025.
These models include DPA-2.4-7M~\cite{zhang2024dpa2,peng2025lambench}, Orb-v3~\cite{rhodes2025orbv3}, MACE-MPA-0~\cite{batatia2023foundation}, SevenNet-MF-ompa~\cite{kim2024data}, MatterSim-v1-5M~\cite{yang2024mattersim} and GRACE-2L-OAM~\cite{bochkarev2024graph}. 
The recently released UMA models~\cite{wood2025umafamilyuniversalmodels} are not accessible in China~\cite{uma-license}, where the authors’ institutions are based, thus we were unable to include them in our benchmark.
The accuracy of each model is quantified using the LWARMSEs for energy, force and, where applicable, virial predictions, as defined in Eq.~\eqref{eq:LWARMSE}.
These metrics are evaluated across all datasets (denoted by ``ALL'') as well as the datasets within the {catalysis}, {inorganic materials}, and {molecules} domains (see Figure~\ref{fig:fftask}).
Detailed error statistics for each dataset are provided in Table~\ref{tab:ffperformance}.
A lower error indicates that the model would either provide superior accuracy as a standalone PES model or require less data for fine-tuning in downstream applications.

The multi-task pretrained DPA-3.1-3M model exhibits state-of-the-art performance across all tasks, achieving overall LWARMSEs of 6.6 meV/atom for energy and 130.6 meV/Å for forces. 
This result underscores the efficacy of multi-task pretraining and the robustness of the DPA-3 architecture in LAM implementations.
Among single-task trained models, Orb-v3 demonstrates competitive performance with average LWARMSEs of 8.7 meV/atom for energy and 171.0 meV/Å for forces, particularly excelling in the Inorganic Materials domain. 
Notably, SevenNet-MF-ompa achieves a lower virial LWARMSE than DPA-3.1-3M, with values of 92.4 meV/atom compared to 98.1 meV/atom, as shown in Table~\ref{tab:ffperformance}.
Although models like Orb-v3 and SevenNet-MF-ompa exhibit comparable or even superior accuracy in certain metrics, they require significantly more parameters than DPA-3.1-3M, with parameter counts of 25.5M and 25.7M versus 3.26M, respectively.

Additionally, most LAMs were trained on datasets labeled exclusively with the PBE XC functional, which does not correspond to the labeling functional used in some downstream test datasets. 
For instance, ANI-1x was originally labeled at the $\omega$B97X/6-31G* level, and AIMD-Chig at the M06-2X/6-31G* level. 
To mitigate the impact of XC functional mismatch on test error, we have relabeled these datasets using the PBE XC functional. 
It is noted that discrepancies in basis sets and pseudopotentials between the training data and downstream tests are not aligned in our evaluation. 
This misalignment primarily results from the significant challenge of reconciling projected augmented wavefunction (PAW) DFT calculations in periodic boundary systems with Gaussian basis set DFT calculations in open boundary systems.
We have chosen not to resolve these mismatches because they have significantly less impact on the test errors than the discrepancies in XC functionals.
For the multi-task trained models DPA-3.1-3M and SevenNet-MF-ompa, we employ the OMat24 dataset encoding during inference, labeling the results as DPA-3.1-3M (OMat) and SevenNet-MF-ompa (OMat), respectively, in Fig.~\ref{fig:fftask} and Tab.~\ref{tab:ffperformance}.

\recheck{When comparing LAM performance, note that training datasets differ across models. 
DPA-3.1-3M and DPA-2.4-7M are multitask-pretrained on OpenLAM-v1, a collection spanning catalysis, inorganic materials, and molecules (see Refs.~\cite{openlam-data-v1-web, peng2025lambench} for details). 
By contrast, the other LAMs are trained primarily on inorganic-materials datasets (see the caption of Fig.~\ref{fig:fftask} for a summary). 
Accordingly, the benchmark in Fig.~\ref{fig:fftask} and Table~\ref{tab:ffperformance} is in-domain but out-of-distribution for DPA-3.1-3M and DPA-2.4-7M, and out-of-domain and out-of-distribution for the other models; hence a systematic domain-wise comparison is warranted. 
Out-of-domain performance is also relevant because several baselines have demonstrated generalization to catalysis and molecular tasks~\cite{batatia2023foundation,rhodes2025orbv3}. From Fig.~\ref{fig:fftask}(a)–(c) and the domain-wise LWARMSE in Table~\ref{tab:ffperformance}, DPA-3.1-3M ranks first in the catalysis and molecules domains, while in inorganic materials it ranks third, slightly behind Orb-v3 and SevenNet-MF-ompa.}

\recheck{
We further assess inorganic-materials generalization using a finetuned variant, DPA-3.1-3M-FT, in which the OMat24 encoding is finetuned on MPTrj and sAlex to align the XC functional. 
On the Matbench Discovery benchmark (Table~\ref{table:sm:noncom:matbench}), DPA-3.1-3M-FT attains the lowest RMSE and the highest $R^2$; its F1 score ranks behind eSEN-30M-OAM, Orb-v3, and SevenNet-MF-ompa, but exceeds GRACE-2L-OAM, MatterSim-v1-5M, and MACE-MPA-0. 
This trend closely matches the LWARMSE ordering for inorganic materials in Table~\ref{tab:ffperformance}, supporting the reliability of our benchmark.}
\recheck{In Supplementary Information Table~\ref{table:sm:spice:ft}, we fine-tuned DPA-3.1-3M on SPICE–MACE–OFF with the SPICE2 dataset encoding and compared it to task-specific DPA3 variants (L12, L24) reported in Fig.~\ref{fig:result}. 
The fine-tuned DPA-3.1-3M achieves comparable accuracy to the DPA3-L12 model, albeit with a slightly larger parameter count (3.3M vs. 2.5M). 
However, it performs less favorably compared to DPA3-L24, which has approximately 50\% more parameters.
The similar performance between the fine-tuned DPA-3.1-3M and the smaller DPA3-L12 model may be due to the fact that DPA-3.1-3M was effectively pretrained on SPICE2 for only 60 epochs, significantly fewer than the 400 epochs used to train DPA3-L12 and DPA3-L24.
}

In contrast to previous works~\cite{shoghi2023molecules,zhang2024dpa2}, which employ separate fitting networks for each dataset during multi-task training, our method utilizes a unified dataset encoding scheme as defined in Eq.~\eqref{eq:atomic-energy}. 
This design ensures that the complexity of the model remains independent of the number of datasets.
To evaluate the scalability and effectiveness of this approach, we conducted a comparative analysis using the OpenLAM-v1 dataset collection. 
We applied the same model architecture as DPA-3.1-3M, with a reduced number of training steps for demonstration purposes, and compared the unified dataset encoding method with the separate fitting strategy\recheck{, see Sec.~\ref{sec:sm:encoding} and Table~\ref{tab:sepshare} for details}.
Generalizability results on force field prediction tasks indicate that the dataset encoding approach achieves comparable
performance across all domains
while requiring significantly fewer model parameters (3.26M compared to 7.69M) than separate fitting networks.
These findings demonstrate the method's advantage as a scalable and efficient solution for LAM development amid increasing dataset diversity.

\section{Discussion}
\label{sec:discussion}

In this work, we introduce the DPA3 model architecture, specifically designed to meet the requirements of large atomistic models (LAMs). 
The DPA3 model adheres to the physical principles inherent in universal potential energy surfaces (PES), exhibiting scaling laws and maintaining a constant parameter scale that remains independent of the number of tasks in a multi-task training scheme. 
The performance of the DPA3 model is demonstrated through two benchmarking approaches.
Firstly, the DPA3 model is assessed in standard potential energy fitting tasks that are well-established in the literature.
Its accuracy is assessed across five benchmark cases spanning diverse research domains, including molecular systems, bulk materials, surface and cluster catalysis, two-dimensional materials, and battery materials.
The model demonstrates superior performance in the majority of test cases.
Secondly, the model is evaluated as a LAM, specifically the DPA-3.1-3M model trained on the OpenLAM-v1 dataset under a multi-task scheme. 
The DPA-3.1-3M model showcases exceptional generalizability in zero-shot force field tasks across 12 downstream datasets spanning distinct scientific domains, indicating its ability to achieve high accuracy as an out-of-the-box potential energy model for scientific inquiries, or requiring minimal data when fine-tuned for downstream tasks.
The DPA3 model is characterized by its smoothness and conservativeness, indicating that higher accuracy in force field generalizability typically correlates with enhanced performance in property calculation tasks~\cite{fu2025learning}.

The pathway to enhancing the generalizability of the DPA3 model as a LAM is evident. 
According to the scaling law, increasing the model size can be achieved through optimizing the implementation or by integrating model parallelization mechanisms.
Additionally, expanding the scale of training data by incorporating datasets from a diverse range of application domains will further enhance the model's generalizability.
For instance, integrating the recently published OMol25 dataset~\cite{levine2025open}, which comprises 102 million configurations, into the training of the DPA3 model is expected to remarkably enhance its generalizability, especially within molecular systems.

It is noteworthy that the DPA3 model currently employs invariant features within the LiGS framework.
Its performance is observed to be suboptimal compared to equivariant models such as MACE when tested on datasets generated by AIMD simulations of small molecules, including the rMD17 dataset~\cite{christensen2020role}, 3BPA dataset~\cite{kovacs2021linear} and acetylacetone dataset~\cite{batatia2022design}.
The potential advantages of incorporating equivariant features to enhance generalizability in these molecular systems remain uncertain and warrant further investigation.

\section{Methods}
\subsection{Datasets}
\label{sec:datasets}
Here, we elaborate on all the benchmark datasets used in Section~\ref{sec:bench}, which span both small molecular systems (e.g. SPICE-MACE-OFF) and material systems (e.g. Water, Zeolite, MPtrj, etc.). 
We used the same dataset splits as prior studies where available. 
For datasets requiring additional processing or splitting, the scripts and processed data have been made publicly available~\cite{zonodo}.



\paragraph{The SPICE-MACE-OFF~\cite{kovacs2023mace}.} This small-molecule dataset, developed for training the MACE-OFF23 model~\cite{kovacs2023mace}, combines 85\% of the SPICE~\cite{eastman2023spice} v1 subset (neutral molecules containing H, C, N, O, F, P, S, Cl, Br, and I) with geometries from classical MD simulations at 300K and 500K for diverse conformations ($\leq$ 50 atoms). 
It extends to larger systems through GFN2-xTB~\cite{bannwarth2019gfn2} MD-generated 50–90 atom molecules (QMugs-derived~\cite{isert2022qmugs}) and water clusters (up to 50 molecules) from liquid water MD trajectories. 
All energies and forces were computed at the $\omega$B97M-D3(BJ)/def2-TZVPPD~\cite{najibi2018nonlocal, weigend2005balanced, rappoport2010property, grimme2011effect, grimme2010consistent} level using PSI4~\cite{smith2020psi4}. Data partitioning follows molecule-level splitting, 95\% training/validation (951,005 structures) and 5\% testing (50,195 structures), to prevent conformation overlap. 
Additional test cases included COMP6 tripeptides~\cite{smith2018less} recomputed at the SPICE level of theory; however, due to the unavailability of these tripeptides in the test set, accuracy comparisons were limited to other molecular systems.

\paragraph{The TorsionNet-500~\cite{rai2022torsionnet}.} This dataset comprises 500 drug-like molecules (H, C, N, O, F, S, Cl) with 24 conformations per molecule, generated by rotating a single bond in 15° increments. 
These structures were initially optimized at the B3LYP/6-31G(d) DFT level and span diverse pharmaceutical chemical space. 

\paragraph{Liquid Water and Ice Dynamics~\cite{zhang2018deep}.} This dataset consists of reference structures for training and validating machine learning potentials (MLPs), encompassing liquid water and ice Ih configurations generated from classical AIMD and PI-AIMD simulations under varying thermodynamic conditions (1-2.13 kbar, 238-330 K) using the PBE0-TS functional. 
It contains 140,000 structures partitioned into 100,000 liquid water configurations (64 \ce{H2O} molecules) and 40,000 ice Ih configurations (96 \ce{H2O} molecules), including three distinct ice phases simulated at PI-AIMD and classical AIMD levels with different pressure-temperature combinations. 
We uniformly sampled <0.1\% of the data (133 structures) for training consistent with previous work~\cite{batzner20223}, with 50 frames allocated for validation and all remaining data reserved for testing. 

\paragraph{The Formate Decomposition on Cu~\cite{batzner20223}.} This dataset consists of configurations characterizing the decomposition process of formate on Cu(110), focusing on C-H bond cleavage. 
It includes initial states (monodentate/bidentate formate), intermediate configurations, and final states (H ad-atom with gas-phase \ce{CO2}). 
The Nudged Elastic Band (NEB) method generated reaction pathways, followed by 12 \textit{ab initio} molecular dynamics (AIMD) simulations using the CP2K~\cite{kuhne2020cp2k} code. 
These simulations produced 6,855 DFT structures with a 0.5 fs time step over 500-step trajectories, capturing dynamic evolution across reaction coordinates. 
The dataset provides atomistic-scale insights into catalytic decomposition mechanisms through systematically sampled configurations. 
The full dataset was partitioned into training (2,500 structures), validation (250 structures), and test (remaining 4,105 structures) sets via uniform random sampling.

\paragraph{The Defected Bilayer Graphene~\cite{ying2024effect}.} This dataset consists of reference structures designed to train and validate MLPs, encompassing three bilayer systems: V0V0 (pristine), V0V1 (single vacancy on the top layer), and V1V1 (single vacancy in both layers). 
The dataset includes single-point DFT (PBE+MBD) energies and atomic forces for configurations with varying interlayer distances, stacking modes, and manual deformations, supplemented by snapshot configurations from classical and DFT-based molecular dynamics (MD) simulations at different temperatures. 
The data were partitioned into training (3,988 structures), validation (4,467 structures), and test (200 structures) sets. 
The splitting was based on farthest point sampling (FPS), by performing principal components analysis (PCA) and choosing sufficiently distant random points for different sets to ensure representative sampling. 

\paragraph{The Zeolite Dataset~\cite{yin2025alphanet}.} This dataset comprises 16 different types of zeolite relevant to catalysis, adsorption, and separation applications. 
It contains atomic trajectories generated through AIMD simulations at 2000 K using VASP~\cite{kresse1996efficiency,kresse1996efficient}, with 80,000 snapshots per zeolite providing calculated energies and atomic forces. 
The dataset is partitioned into training (48,000 structures), validation (16,000), and test (16,000) sets using a random 6:2:2 split ratio. 
This standardized division ensures systematic evaluation of MLPs while maintaining consistency across computational frameworks.

\paragraph{The DPA2 Test Sets~\cite{zhang2024dpa2}.}
This composite dataset comprises 18 specialized sub-datasets (totaling 5,119,379 structures: 4,045,094 training, 1,074,285 test) for pre-training the DPA2 model, integrating \textbf{domain-specific collections} including metallic alloys (Alloy), cathode materials (Cathode-P), semiconductors (SemiCond-P), drug-like molecules (Drug), catalytic reaction trajectories (OC2M) and etc., alongside with specialized systems such as \ce{H2O} configurations, metallic materials (Sn, AgAu, AlMgCu), and n-dodecane pyrolysis (C12H26). 
The '-P' suffix indicates datasets reformulated with pretraining splits in the DPA2 paper (see Ref.~\cite{zhang2024dpa2} for details). 
Aggregated from the DeepModeling community~\cite{deepmodeling-community} and external sources, these 73-element datasets employ calculations from various DFT softwares like the VASP~\cite{kresse1996efficiency,kresse1996efficient}, Gaussian~\cite{frisch2016gaussian}, and ABACUS~\cite{chen2010systematically, li2016large} for multi-task training. 
Following Ref.~\cite{zhang2024dpa2}, we conduct per-dataset training with averaged error metrics for fair benchmarking.

\paragraph{Materials Project Trajectory Dataset (MPtrj)~\cite{deng2023chgnet}.} This dataset provides a comprehensive collection of DFT calculations for $\sim$146,000 inorganic materials spanning 89 elements, derived from 1.37 million structural relaxation and static calculation tasks in the Materials Project Database~\cite{jain2013commentary}. 
Calculations employ generalized gradient approximation (GGA) or GGA+U exchange-correlation methods, generating 1.58 million atomic configurations with associated energies, magnetic moments (7.94 million), atomic forces (49.3 million), and stress tensors (14.22 million). 
By systematically sampling trajectories from relaxation pathways and equilibrium states, the dataset captures the potential energy surfaces from diverse regions of inorganic materials, serving as a foundational resource for training and benchmarking machine learning models in computational materials science. 
We employ this dataset to train models and test their ability to predict ground-state (0 K) thermodynamic stability through geometry optimization and energy prediction within the Matbench Discovery Benchmark~\cite{riebesell2023matbench}.

\subsection{Line graph transform} \label{sec:ligs-explain}

\begin{figure}
    \centering
    \includegraphics[width=0.9\linewidth]{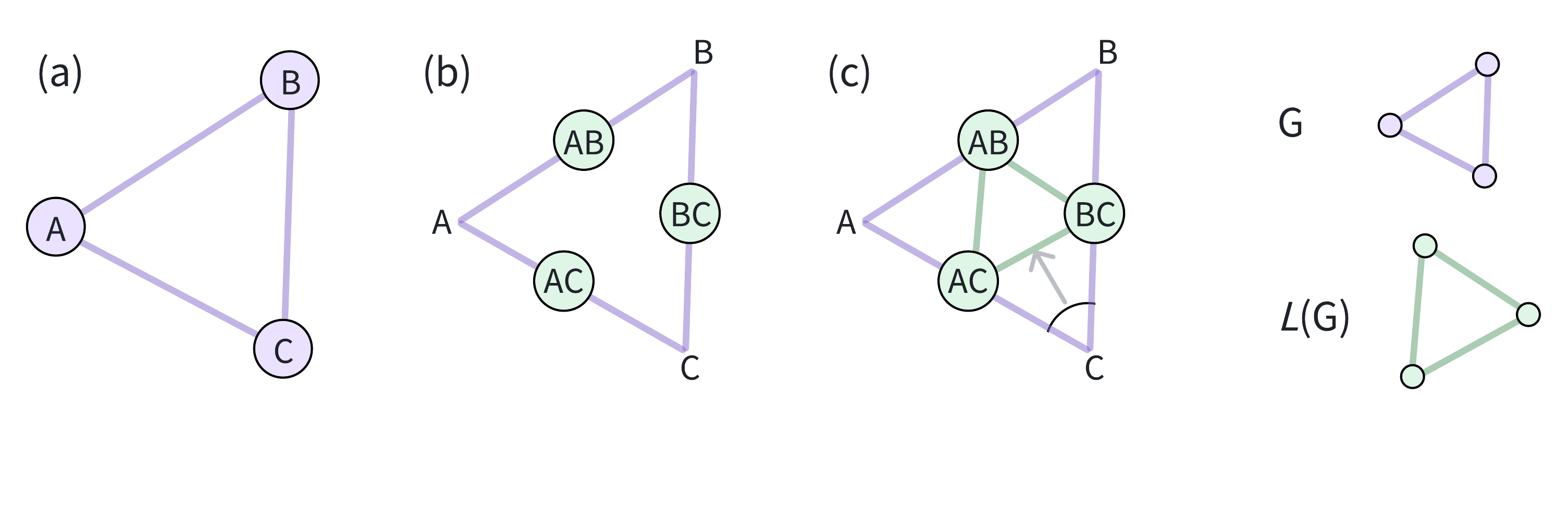}
    \caption{\recheck{Graphic illustration of line graph transform. 
    (a) The initial graph, denoted as \( G \), is depicted in purple and comprises three vertices (atoms), labeled A, B, and C.
    (b)  The vertices of the transformed line graph, designated as \( \mathcal{L}(G) \), are illustrated in green and labeled AB, AC, and BC, corresponding to the edges AB, AC, and BC in \( G \).
    (c) Edges in \( \mathcal{L}(G) \), shown in green, are established between vertices when their corresponding edges in \( G \) share a common vertex. 
    The correspondence between an edge in \( \mathcal{L}(G) \), specifically AB-BC, and the angle ACB is highlighted by a gray arrow.
    }}
    \label{fig:ligs-explain}
\end{figure}

\recheck{
The line graph transformation is depicted in Fig.~\ref{fig:ligs-explain}. 
Consider three atoms, labeled A, B, and C, which are treated as vertices in a graph. 
These atoms are sufficiently proximate such that each pair is considered neighbors.
A fully connected graph \( G \) is constructed by connecting the vertex pairs AB, BC, and AC with edges, as shown in Fig.~\ref{fig:ligs-explain}(a). 
The line graph transformation initially assigns the vertices of \( \mathcal{L}(G) \) to each edge of the graph \( G \), as illustrated in Fig.~\ref{fig:ligs-explain}(b). 
Consequently, each vertex in \( \mathcal{L}(G) \) corresponds to a neighboring atom pair.
If two edges in \( G \) share a common vertex, their corresponding vertices in \( \mathcal{L}(G) \) are connected by an edge, as demonstrated in Fig.~\ref{fig:ligs-explain}(c).
For instance, the edges AC and BC in \( G \) share the common vertex C, resulting in the nodes AC and BC in \( \mathcal{L}(G) \) being linked by an edge, denoted as AC-BC.
This edge directly corresponds to the angle A-C-B in the original graph \( G \), as highlighted by the gray arrow in Fig.~\ref{fig:ligs-explain}(c).
}

\subsection{DPA3 model architecture}\label{sec:dpa3-arch}

In this study, we investigate a system composed of \( N \) atoms, where the atomic numbers are represented by the list \( \mathcal{Z} = \left(Z_1, \dots, Z_i, \dots, Z_N \right) \) and the atomic positions are denoted by \( \mathcal{R} = \left(\bm{r}_1, \dots, \bm{r}_i, \dots, \bm{r}_N \right) \). 
The PES of the system, indicated by \( E = E(\mathcal{Z}, \mathcal{R}) \), is postulated to be decomposable into a sum of atomic contributions \( E_i \), such that \( E = \sum_i E_i \).
The atomic force \( {F}_i \) exerted on atom \( i \) and the virial tensor \( \Xi_{pq} \) are defined as the negative gradients of the total energy with respect to the atomic coordinates and the simulation cell vectors, respectively:
\begin{align}
\label{eq:fv}
{F}_{i}=-\nabla_{\bm{r}_i} E, \quad
\Xi_{pq}&=-\sum_{r} \frac{\partial E}{\partial \textbf{h}_{rp}} \textbf{h}_{rq}.
\end{align}
In Eq.~\eqref{eq:fv} $\Xi_{pq}$ corresponds to the $pq$ component of the virial tensor, and $\textbf{h}_{pq}$ yields the $q$-th component of the $p$-th cell vector.

The DPA3 model constructs the atomic energy contribution as follows:
\begin{align}
    \label{eq:atomic-energy}
    E_i = \mathcal F\left( v_i^{(1,L)}, c(\mathcal D_m) \right) + e_{m}(Z_i),
\end{align}
In this formulation, \(\mathcal{F}\) denotes the fitting network, with \(v_i^{(1,L)}\) representing the vertex feature of graph \(G^{(1)}\) after \(L\) layers of updates.
This vertex feature functions as a descriptor of the atomic environment surrounding atom \(i\).
Furthermore, \(c(\mathcal{D}_m)\) refers to the encoding of the dataset \(\mathcal{D}_m\).
The energy bias, \(e_{m}(Z_i)\), is obtained through least squares fitting to the training energy of dataset $\mathcal D_m$, utilizing the chemical formula.
This approach effectively mitigates the arbitrariness associated with selecting the energy zero point across different DFT calculations.

The DPA3 model evolves vertex and edge features through a recursive formulation applied to each graph within the LiGS.
Let \(v_\alpha^{(k,l)}\) and \(e_{\alpha\beta}^{(k,l)}\) denote the vertex and edge features of graph \(k\) after \(l\) layers, respectively.
The features of the $l+1$ layer are initialized by the features of the previous layer, i.e., \( v_\alpha^{(k,l+1)} \gets v_\alpha^{(k,l)} \) and \( e_{\alpha\beta}^{(k,l+1)} \gets e_{\alpha\beta}^{(k,l)} \). 
The features are subsequently refined using a residual update mechanism:
\begin{align}\label{eq:nupdate}
    & v_\alpha^{(k,l+1)} \gets  v_\alpha^{(k,l+1)} + \delta_c^{(k,l)} u_\alpha^{(k,l)}, \\\label{eq:eupdate}
    & e_{\alpha\beta}^{(k,l+1)} \gets  e_{\alpha\beta}^{(k,l+1)} + \delta_s^{(k,l)}  m_{s,\alpha\beta}^{(k,l)},
\end{align}
where \(\delta_c^{(k,l)}\) and \(\delta_s^{(k,l)}\) represent trainable step sizes for the updates. The term \( u_\alpha^{(k,l)} \) indicates the vertex feature update, defined as a convolution of messages of all the edges connecting to the vertex. 
\( m_{s,\alpha\beta}^{(k,l)} \) denotes the self-message used in the edge feature update. 
These updates are defined as follows:
\begin{align}\label{eq:defu}
    &u_\alpha^{(k,l)} = \phi_u \left(
    N_m^{-{\alpha_k}}\sum_{\beta\in \mathcal E^{(k)}(\alpha)} w_{\alpha\beta}^k\, m_{c,\alpha\beta}^{(k,l)}
    \right),\\\label{eq:cmsg}
    & m^{(k,l)}_{c,\alpha\beta} = \phi_c( v_\alpha^{(k,l)}, v_\beta^{(k,l)}, e_{\alpha\beta}^{(k,l)})
    \\\label{eq:smsg}
    & m^{(k,l)}_{s,\alpha\beta} = \phi_s( v_\alpha^{(k,l)}, v_\beta^{(k,l)}, e_{\alpha\beta}^{(k,l)})
\end{align}
where \(\phi_u\) may be either a multi-layer perceptron (MLP) or an identity mapping, and \(\phi_c\) and \(\phi_s\) are MLPs. 
In Eq.~\eqref{eq:defu}, \(\mathcal{E}^{(k)}(\alpha)\) represents the set of all edges connected to vertex \(\alpha\) in the graph \(G^{(k)}\).
The normalization factor \(N_m^{-{\alpha_k}}\) involves \(N_m\), which estimates the maximum possible number of neighbors, and a power factor \(\alpha_k > 0\).
Notably, the vertex feature \( v^{(k,l)} \) is identical to the edge feature \( e^{(k-1,l)} \) from the preceding graph in the LiGS.
Consequently, we avoid maintaining redundant information regarding vertex features for any graph \(G^{(k)}\) where \(k > 1\). 
Instead, the vertex feature updates \(u_\alpha^{(k,l)}\) are transferred to graph \(G^{(k-1)}\) to update the edge features  \( e^{(k-1,l)} \).
Subsequently, the updated edge features \( e^{(k-1,l+1)} \) of graph \(G^{(k-1)}\) determine the vertex feature \(v^{(k,l+1)}\) of graph \(G^{(k)}\), as depicted in Fig.~\ref{fig:DPA-3-arch}(e).

In Eq.~\eqref{eq:defu}, the message \(m^{(k,l)}_{c,\alpha\beta}\) is appropriately smoothed by incorporating a prefactor \(w_{\alpha\beta}^k\) to ensure that it gradually diminish as the distance between the atoms approaches the cut-off radius at different orders $k$. 
Specifically, in $G^{(1)}$, we define a smooth switch function $s^k(r_{ij})$  that exponentially decays from 1 to 0 as the distance $r_{ij}$ defined  increases from 0 to a cutoff radius $r_{c}^k$. 
The switch function is defined as:
\begin{align}\label{eq:sw}
    s^k\left(r_{ij}\right)=  
\begin{cases}
\exp(-\exp( C (r_{ij} - r_{cs}^k)/r_{cs}^k )) & \text{if } 0 < r_{ij} \leq r_{c}^k,  \\ 
0 & \text{if } r_{ij} > r_{c}^k,
\end{cases}
\end{align}
where $r_{cs}^k$ is a manually adjustable smoothing factor and $C$ is a tunable hyper-parameter controlling the decay rate. 
Typically, for $k=1$, we set $w_{ij}^1 = s^1(r_{ij})$, with $C = 20$, $r_c^1 = 6.0\text{\AA}$, and $r_{cs}^1 = 5.3\text{\AA}$.
For higher-order terms ($k>1$), a similar approach is used, but the prefactor $w_{\alpha\beta}^k$ is defined as the product of multiple switch functions, each acting on the distances defined within $G^{(1)}$, ensuring overall smooth diminishing. 
In graph \( G^{(2)} \), the edge \(\alpha\beta\) is represented by the angle formed between two atomic bonds that share a common vertex, specifically \((ij)(im)\). 
Thus,
\begin{align}
w_{\alpha\beta}^2 = w_{(ij)(im)}^2 &= s^2(r_{ij}) \times s^2(r_{im}).
\end{align}
In graph $G^{(3)}$, the edge \(\alpha\beta\) is represented by the dihedral formed between two angles that share a common bond, specifically \((ijm)(ijn)\).
\begin{align}
w_{\alpha\beta}^3  = w_{(ijm)(ijn)}^3 &= s^3(r_{ij}) \times s^3(r_{im}) \times s^3(r_{in}).
\end{align}

For the initial graph \( G^{(1)} \), vertex features are iteratively refined by incorporating additional self-message and symmetrization transformation terms as follows (see Fig.~\ref{fig:DPA-3-arch}~(d)):
\begin{align}\label{eq:n0self}
    v_i^{(1,l+1)} \gets  v_i^{(1,l+1)} + \delta_{s,0}^{(1,l)}  \phi_{s,0} (v_i^{(1,l)})+ \delta_{s,1}^{(1,l)}  \phi_{s,1} (\tilde v_i^{(1,l)}).
\end{align}
In this equation, \(\phi_{s,0}\) and \(\phi_{s,1}\) are the MLPs responsible for processing the self-message and the symmetrization-transformed intermediate features \(\tilde{v}_i^{(1,l)}\), respectively.
The intermediate representation $\tilde v_i^{(1,l)}$ is defined by
\begin{align}\label{eq:symm}
   \tilde v_i^{(1,l)} = 
   \mathrm{concat}
   \bigg(
   \mathrm{symm}\big( v_j^{(1,l)}, h_{ij}\big), \, 
   \mathrm{symm}\big( e_{ij}^{(1,l)}, h_{ij}\big)
   \bigg).
\end{align}
Here, \(h_{ij}\) is a three-component vector given by \(h_{ij} = \frac{s^1(r_{ij})}{(r_{ij})^2} \times (x_{ij}, y_{ij}, z_{ij})\), where \((x_{ij}, y_{ij}, z_{ij})\) represent the Cartesian components of the vector \(\bm r_{ij} = \bm r_i - \bm r_j\), \(r_{ij} = \vert \bm r_{ij}\vert\) and $s^1(r_{ij})$ is the smooth switch function defined in Eq.~\eqref{eq:sw} with $k=1$.
The \(\mathrm{symm}\) operator, as introduced by Zhang et al.~\cite{zhang2018deep}, is defined generally as \(\mathrm{symm}(a_j, b_j)\), with \(a_j\) and \(b_j\) being neighbor-indexed vectors. 
It is assumed that \(a_j\) is rotationally invariant, whereas \(b_j\) is not, although their inner product maintains rotational invariance.
The symmetrization operator is mathematically represented by:
\begin{align}\label{eq:symm-2}
    &\mathrm{symm}(a_j, b_j) = 
    \underset{ps}{\mathrm{flatten}}\Big(
    \sum_q \psi_{pq} \, \psi^<_{sq}
    \Big),\\\label{eq:symm-0}
    &\psi_{pq}  = \frac{1}{\sqrt{N_m}}
    \sum_{j\in N_{r_c^1}(i)}
    w^1_{ij}\, a_{j,p}\, b_{j,q}, \\\label{eq:symm-1}
    &\psi^<_{pq} = \underset{p}{\mathrm{split}}(\psi_{pq}).
\end{align}
In Eq.~\eqref{eq:symm-2}, the output matrix is flattened into a vector along the dimensions \(p\) and \(s\). 
Eq.~\eqref{eq:symm-0} involves a summation over neighboring indices \(j\), ensuring that the matrix \(\psi_{pq}\) remains permutationally invariant. 
Eq.~\eqref{eq:symm-1} describes the splitting of the matrix \(\psi_{pq}\) along the dimension \(p\), where the first certain number of elements is taken and denoted by \(\psi^<_{pq}\). 
It can be demonstrated that the symmetrization operator is invariant under rotational operations and permutations of atoms sharing the same atomic number, as detailed in Ref.~\cite{zhang2018end}.

The vertex features of graph \(G^{(1)}\) correspond to attributes associated with the atoms within the system and are initialized as follows:
\begin{align}
\label{eq:init-feat-v1}
v_\alpha^{(1,0)} = v_i^{(1,0)} = \operatorname{one\_hot}(Z_i),
\end{align}
where $\operatorname{one\_hot}$ denotes the one-hot encoding of atomic number $Z_i$.
The edge features of the graph represent neighboring atom pairs and are initialized through:
\begin{align}
\label{eq:init-feat-e1}
e_{\alpha\beta}^{(1,0)} =e_{ij}^{(1,0)} = \phi^{(1)}_e({r}_{ij}), 
\end{align}
where \(\phi^{(1)}_e\) is an MLP designed to embed the interatomic distances between pairs of atoms.
In graph \(G^{(2)}\), the edge \(\alpha\beta\) corresponds to an angle formed by two bonds \((ij)\) and \((im)\) sharing a common vertex. 
The edge features $e_{\alpha\beta}^{(2,0)}$ are initialized using:
\begin{align}
\label{eq:init-feat-e2}
    e_{\alpha\beta}^{(2,0)} = e_{(ij)(im)}^{(2,0)}  = \phi^{(2)}_e(\cos(\theta_{ijm})) , 
\end{align}
where \(\phi^{(2)}_e\) represents the embedding network, and \(\theta_{ijm}\) is the angle formed by the atom pairs \(ij\) and \(im\).
For graph \(G^{(3)}\), the edge \(\alpha\beta\) corresponds to a dihedral formed by two angles \((ijm)\) and \((ijn)\) sharing a common bond $(ij)$. The edge features are initialized by:
\begin{align}
\label{eq:init-feat-e3}
    e_{\alpha\beta}^{(3,0)}  = 
    e_{(ijm)(ijn)}^{(3,0)}  = \phi^{(3)}_e(\cos(\eta_{mijn})) , 
\end{align}
where \(\phi^{(3)}_e\) denotes the embedding network, and \(\eta_{mijn}\) is the dihedral angle involving the angle objects \(jim\) and \(jin\). 
These initializations ensure that each graph layer is equipped with features pertinent to its respective structural relations.

To demonstrate the invariance of the DPA3 model under translation, rotation, and permutation operations, it is sufficient to establish that the vertex feature of the final layer, $v_i^{(1,L)}$, remains equivariant under permutation operations and invariant under translational and rotational operations. 

For translational and rotational symmetries, it is sufficient to confirm that both feature initializations and layer-wise updates maintain these symmetries. 
The vertex and edge features of the graphs are initialized according to Eqs.~\eqref{eq:init-feat-v1}--\eqref{eq:init-feat-e3}, which consider atomic numbers, interatomic distances, angles, and dihedrals - all of which are invariant under translational and rotational transformations.
Moreover, since the inputs to the messages in Eqs.~\eqref{eq:cmsg} and \eqref{eq:smsg} are translationally and rotationally invariant, the messages themselves are also invariant under these operations. Consequently, the layer-wise feature updates preserve translational and rotational invariance.

To examine the equivariance concerning permutational operations over atoms of identical chemical species, we initially note that the vertex set, edge set, and connections remain invariant under these operations across all graphs in the LiGS.
Next, we consider feature initialization. 
The initial vertex features in graph \(G^{(1)}\), as described by Eq.~\eqref{eq:init-feat-v1}, are equivariant. Although the order of the edge features \(e_{\alpha\beta}^{(k,0)}\) for any \(k \geq 1\) may change, these altered features can be mapped back to the original set via a permutation.
During the feature update process, the messages defined in Eqs.~\eqref{eq:cmsg} and \eqref{eq:smsg} exhibit the same permutation pattern as the edges. 
Given that the connections \(\mathcal{E}^{(k)}\) within any graph \(G^{(k)}\) remain unchanged, the convolution in Eq.~\eqref{eq:defu} is aligned with the vertex permutation. 
This alignment is attributable to the fact that the order of neighbors \(\beta\) is irrelevant due to the summation involved.
Consequently, according to Eqs.~\eqref{eq:nupdate} and \eqref{eq:eupdate}, both vertex and edge updates maintain the same permutation pattern as their respective features.
Since the vertex features of the graph $G^{(1)}$ are permuted in the same manner as the atomic permutation, the permutational equivariance of the final layer \(v_i^{(1,L)}\) is confirmed.

\section{Acknowledgments}
We gratefully acknowledge the support received for this work. 
The work of Han Wang is supported by the National Key R\&D Program of China (Grant No.~2022YFA1004300). 
The work of Jinzhe Zeng was supported by the Advanced Materials - National Science and Technology Major Project (No. 2025ZD0618702), the advanced computing resources provided by the Supercomputing Center of the USTC and the Open Source Supercomputing Center of S-A-I.
The work of Tong Zhu is supported by the National Natural Science Foundation of China (Grants No.~22222303 and No.~22173032).

\section{Data availability}
The OpenLAM-v1 dataset and the pretrained DPA-3.1-3M model are all available on AIS Square (\url{https://www.aissquare.com/datasets/detail?pageType=datasets&name=OpenLAM-TrainingSet-v1&id=308} and \url{https://www.aissquare.com/models/detail?pageType=models&name=DPA-3.1-3M&id=343}), respectively. 
The DPA3 training and inference codes are available in the DeePMD-kit repository (\url{https://github.com/deepmodeling/deepmd-kit}) after version 3.1.0. 
The datasets used to evaluate zero-shot generalization, together with scripts to reproduce the reported error metrics, are available in Ref.~\cite{zonodo}.






\renewcommand{\thefigure}{S-\arabic{figure}}
\setcounter{figure}{0}
\renewcommand{\thetable}{S-\arabic{table}}
\setcounter{table}{0}
\renewcommand{\thesection}{S-\arabic{section}}
\setcounter{section}{0}
\renewcommand{\theequation}{S\arabic{equation}}
\setcounter{equation}{0}

\FloatBarrier 
\begin{center}
    {\LARGE\bfseries Supplementary Information \par}
    \vspace{1em}
\end{center}

\section{Benchmark results}
\subsection{Detailed benchmark tables}
Here we present detailed benchmark results of the DPA3 model across various datasets as discussed in the main text, including SPICE-MACE-OFF (Table~\ref{table:spice}), TorsionNet-500 (Table~\ref{table:torsion}), liquid water and three ice systems (Table~\ref{table:water}), three material datasets from Ref.~\cite{yin2025alphanet} (Table~\ref{table:fcu}) and DPA2 test sets (Table~\ref{table:single_task}). 
We then present detailed zero-shot generalizability results of the DPA-3.1-3M model, compared against several state-of-the-art LAMs, on force field tasks across 12 downstream datasets in Table~\ref{tab:ffperformance}. 
\recheck{
We also show the finetuned results of DPA-3.1-3M on Matbench Discovery (Table~\ref{table:sm:noncom:matbench}) and SPICE-MACE-OFF (Table~\ref{table:sm:spice:ft}) compared with task-specific models only trained on these datasets.
}

\subsection{Correlation for downstream properties}
\label{sec:app:energy-property}

\begin{figure}
    \centering
    \includegraphics[width=1.0\textwidth]{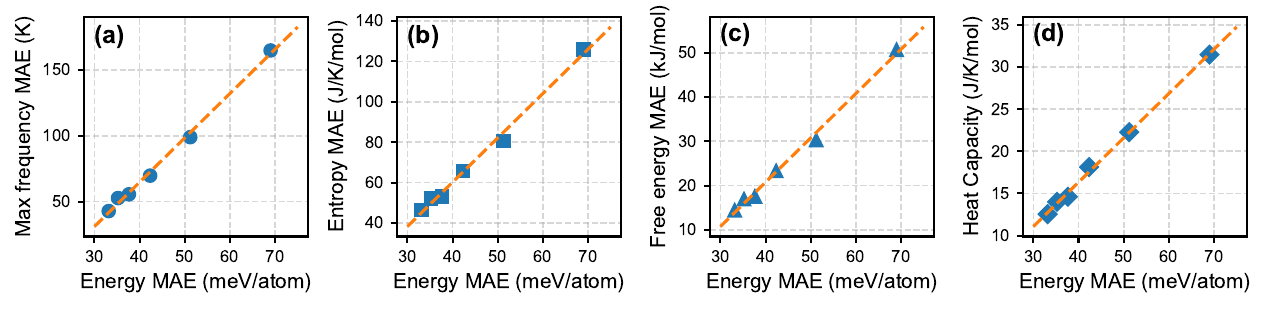}
    \caption{
    \recheck{Correlation between energy MAE and downstream phonon-related property MAEs.
    DPA3 variants trained on MPtrj with varying energy MAE on the WBM test set are evaluated on the MDR benchmark~\cite{loew2025universal} for phonon calculation tasks.
    Shown are MAEs for (a) maximum frequency, (b) entropy, (c) free energy, and (d) heat capacity.}
    }
    \label{fig:sm:phonon}
\end{figure}

\recheck{
We find a correlation between test error of DPA3 models and downstream property prediction—an effect also observed for eSEN~\cite{fu2025learning}. 
To make this dependence explicit, we first trained a set of MPtrj models spanning different energy-accuracy levels and evaluated their energy MAE on the WBM test set. 
To widen the MAE range, we selected DPA3 checkpoints with different parameter counts (L3/L6/L9/L16) and taken at different training stages. 
We then ran calculation tasks on the MDR benchmark~\cite{loew2025universal} to obtain phonon-related properties. 
As shown in Fig.~\ref{fig:sm:phonon}(a-d), the MAEs of four properties (max frequency, entropy, free energy and heat capacity) decrease as the model’s energy MAE decreases, exhibiting a clear approximately linear trend. 
This demonstrates strong correlation between energy accuracy and downstream properties, and indirectly indicates that the scaling law is likewise meaningful for property prediction—i.e., further reductions in energy MAE/RMSE via scaling are expected to yield commensurate improvements in downstream properties.
}

\subsection{Inference efficiency evaluation}
\label{sec:sm:effi}

\begin{figure}
    \centering
    \includegraphics[width=0.5\linewidth]{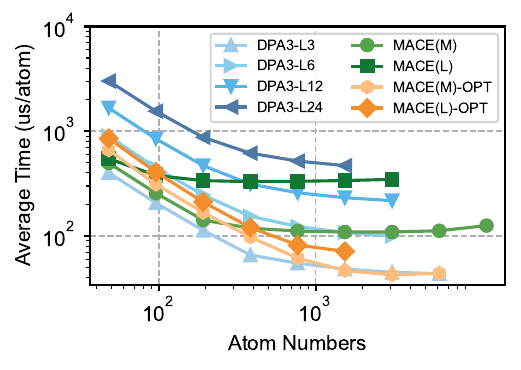}
    \caption{
    \recheck{Inference efficiency of different model architectures.
    Comparison of DPA3 (L3, L6, L12, and L24) and MACE (M and L) on an extensible water system.
    “{-OPT}” denotes optimized MACE implementations using the \texttt{cuda\_mace}~\cite{cuda_mace,kovacs2023mace} package.
    All simulations were performed end-to-end via the ASE calculator on a single 80 GB Nvidia A800 GPU, with system size increased from 48 atoms (16 \ce{H2O} molecules) until out-of-memory.
    }
    }
    \label{fig:sm:effi}
\end{figure}

\recheck{To compare the inference efficiency of different model architectures, we benchmarked DPA3 (L3, L6, L12, and L24), MACE (M and L), and optimized MACE implementations using \texttt{cuda\_mace}~\cite{cuda_mace,kovacs2023mace} with custom CUDA kernels.
To systematically assess how efficiency scales with system size, we employed an extensible water system, increasing the atom count from 48 atoms (16 \ce{H2O} molecules) to up to 12,288 atoms (4096 \ce{H2O} molecules)  while maintaining the equilibrium density.
All experiments were executed end-to-end using the ASE calculator on a single 80 GB Nvidia A800 GPU.
For each model, the system size was increased until an out-of-memory condition occurred, and the average inference time was measured over 100 repetitions.
Figure~\ref{fig:sm:effi} shows the average per-atom inference time (in microseconds) as a function of system size, where the optimized implementations are denoted as MACE(M)-OPT and MACE(L)-OPT.
}

\begin{table}
  \scriptsize
  \caption{Test MAE on SPICE-MACE-OFF dataset. Energy (E) MAE is in meV/atom and force (F) MAE is in meV/\AA.}
  \label{table:spice}
  \centering
  \begin{threeparttable}
  \begin{spacing}{1.15}
  \begin{tabular}{l
    | R{0.55cm}R{0.55cm}  
    | R{0.55cm}R{0.55cm}  
    | R{0.55cm}R{0.55cm}  
    | R{0.55cm}R{0.55cm}  
    | R{0.55cm}R{0.55cm}  
    | R{0.55cm}R{0.55cm}  
    | R{0.55cm}R{0.55cm}  
    | R{0.55cm}R{0.55cm}  
  }
    \toprule
     & \multicolumn{2}{c|}{MACE(M)} & \multicolumn{2}{c|}{MACE(L)} & \multicolumn{2}{c|}{EScAIP\tnote{a}} & \multicolumn{2}{c|}{eSEN} & \multicolumn{2}{c|}{DPA3-L3} & \multicolumn{2}{c|}{DPA3-L6} & \multicolumn{2}{c|}{DPA3-L12} & \multicolumn{2}{c}{DPA3-L24} \\\hline
    \textbf{Dataset} & \textbf{E~~~} & \textbf{F~~~} & \textbf{E~~~} & \textbf{F~~~} & \textbf{E~~~} & \textbf{F~~~} & \textbf{E~~~} & \textbf{F~~~} & \textbf{E~~~} & \textbf{F~~~} & \textbf{E~~~} & \textbf{F~~~} & \textbf{E~~~} & \textbf{F~~~} & \textbf{E~~~} & \textbf{F~~~} \\
    PubChem      & 0.91 & 20.57 & 0.88 & 14.75 & 0.53 & \underline{5.86} & \textbf{0.15} & \textbf{4.21} & 0.96 & 20.61 & 0.51 & 15.20 & 0.41 & 12.12 & \underline{0.24} & 8.47 \\
    DES370K M.   & 0.63 &  9.36 & 0.59 &  6.58 & 0.41 &  3.48 & \textbf{0.13} & \textbf{1.24} & 0.53 &  8.57 & 0.35 &  6.37 & 0.31 &  5.25 & \underline{0.18} & \underline{3.15} \\
    DES370K D.   & 0.58 &  9.02 & 0.54 &  6.62 & 0.38 & \underline{2.18} & \textbf{0.15} & \textbf{2.12} & 0.70 &  7.89 & 0.46 &  6.14 & 0.43 &  5.13 & \underline{0.23} & 3.19 \\
    Dipeptides   & 0.52 & 14.27 & 0.42 & 10.19 & 0.31 &  5.21 & \textbf{0.07} & \textbf{2.00} & 0.58 & 12.26 & 0.26 &  9.16 & 0.24 &  7.55 & \underline{0.13} & \underline{4.81} \\
    Sol. AA      & 1.21 & 23.26 & 0.98 & 19.43 & 0.61 & 11.52 & \textbf{0.25} & \textbf{3.68} & 0.76 & 20.38 & 0.60 & 15.48 & 0.38 & 12.69 & \underline{0.31} & \underline{8.77} \\
    Water        & 0.76 & 15.27 & 0.83 & 13.57 & 0.72 & 10.31 & \textbf{0.15} & \textbf{2.50} & 0.65 & 14.57 & 0.59 & 11.69 & 0.47 &  9.83 & \underline{0.32} & \underline{6.89} \\
    QMugs        & 0.69 & 23.58 & 0.54 & 16.93 & 0.41 &  8.74 & \textbf{0.12} & \textbf{3.78} & 1.04 & 21.61 & 0.34 & 16.17 & 0.32 & 12.90 & \underline{0.17} & \underline{8.66} \\\hline
    LWAMAE\tnote{b} & 0.73 & 15.42 & 0.65 & 11.66 & 0.46 &  5.87 & \textbf{0.14} & \textbf{2.58} & 0.73 & 14.08 & 0.43 & 10.69 & 0.36 &  8.74 & \underline{0.22} & \underline{5.78} \\\hline
    \textbf{\# Params} & \multicolumn{2}{c|}{2.3 M} & \multicolumn{2}{c|}{6.9 M} & \multicolumn{2}{c|}{45 M} & \multicolumn{2}{c|}{6.5 M} & \multicolumn{2}{c|}{0.9 M} & \multicolumn{2}{c|}{1.3 M} & \multicolumn{2}{c|}{2.5 M} & \multicolumn{2}{c}{4.9 M} \\
    \toprule
  \end{tabular}

  \begin{tablenotes}
    \item[a] \footnotesize Model uses direct-force prediction.
    \item[b] \footnotesize The Logarithmic Weighted Average MAE defined in Eq.~\eqref{eq:LWARMSE}.
  \end{tablenotes}
  \end{spacing}
  \end{threeparttable}
\end{table}

\begin{table}
\caption{Torsional MAE, RMSE and MAEB errors (all in kcal/mol) between MLIP predictions and its reference DFT labels on the TorsionNet-500 dataset, following Ref.~\cite{yang2024ab}.} \label{table:torsion}
\centering
\begin{threeparttable}
\begin{spacing}{1.3}
 \begin{tabular}{l| c | cccc}
  \toprule
   \textbf{Model} & \textbf{Training data} & \textbf{MAE} $\downarrow$ & \textbf{RMSE} $\downarrow$ & \textbf{MAEB\tnote{a}} $\downarrow$ & \textbf{NABH$_{h}$\tnote{b}} $\downarrow$\\
       \hline 
   AIMNet2 & Ref.~\cite{anstine2024aimnet2} & 0.38 & 0.55 & 0.58 & 82\\
   DPA2-Drug   & Ref.~\cite{yang2024ab} & 0.24 & 0.35 & 0.36 &  18\\
   MACE(M) & SPICE-MACE-OFF & 0.19 & 0.28 & 0.31 &  7\\
   MACE(L) & SPICE-MACE-OFF & 0.14 & 0.21 & 0.23 &  4\\
   DPA3-L3 & SPICE-MACE-OFF & 0.17 & 0.25 & 0.26 &  7\\
   DPA3-L6 & SPICE-MACE-OFF & 0.13 & 0.19 & 0.20 &  3\\
   DPA3-L12 & SPICE-MACE-OFF & 0.10 & 0.16 & 0.17 &  \textbf{0}\\
   DPA3-L24 & SPICE-MACE-OFF & \textbf{0.06} & \textbf{0.09} & \textbf{0.09} &  \textbf{0}\\
   Nutmeg(L) & SPICE2 & - & - & 0.20~\cite{eastman2024nutmeg} &  -\\ 
   \toprule
 \end{tabular}
 \begin{tablenotes}
    \item[a] \footnotesize The MAE of the torsional barrier height, defined as the difference between the minimum and the maximum energy points during the torsional rotation.
    \item[b] \footnotesize The number of molecules (total:$N_{\mathrm{mols}}=500$) for which the model prediction of potential barrier height has an error of more than 1 kcal/mol.
 \end{tablenotes}
 \end{spacing}
 \end{threeparttable}
\end{table}

\begin{table}
  \footnotesize
  \caption{Test RMSE on the liquid water and three ice systems~\cite{zhang2018deep}. 
  Energy (E) RMSE is in meV/\ce{H2O}, force (F) RMSE is in meV/\AA. 
  Except for DeePMD, models are trained on 0.1\% of the training data following Ref.~\cite{batzner20223}.} 
  \label{table:water}
  \centering  
  \begin{threeparttable}
  \begin{spacing}{1.3}
  \begin{tabular}{l | R{0.7cm}R{0.7cm} | R{0.6cm}R{0.6cm} | R{0.6cm}R{0.6cm} | R{0.6cm}R{0.6cm} | R{0.6cm}R{0.6cm} | R{0.6cm}R{0.6cm}}
    \toprule
     & \multicolumn{2}{c|}{\textbf{100\% data}} & \multicolumn{10}{c}{\textbf{0.1\% data}} \\\cline{2-13}
     & \multicolumn{2}{c|}{DeePMD} & \multicolumn{2}{c|}{DeepPot-SE} & \multicolumn{2}{c|}{NequIP} & \multicolumn{2}{c|}{DPA3-L6} & \multicolumn{2}{c|}{DPA3-L12} & \multicolumn{2}{c}{DPA3-L24} \\\hline
    \textbf{System} & \textbf{E~~~} & \textbf{F~~~} & \textbf{E~~~} & \textbf{F~~~} & \textbf{E~~~} & \textbf{F~~~} & \textbf{E~~~} & \textbf{F~~~} & \textbf{E~~~} & \textbf{F~~~} & \textbf{E~~~} & \textbf{F~~~} \\
    Liquid Water  &  \underline{1.0} & \underline{40.4}  &  1.8  &  66.8  &  1.6 & 51.4  &  \textbf{0.7}  &  \textbf{37.3} &  \underline{1.0}  &  41.9  &  \underline{1.0}  &  44.5\\
    Ice Ih (b)  &  \textbf{0.7} & 43.3 &  4.7  &  61.8  &  2.5 & 57.8  &  \underline{1.2}  &  \underline{32.3} &  1.4  &  \textbf{31.4}  &  1.3  &  \underline{32.3}\\
    Ice Ih (c)  &  \textbf{0.7} & 26.8 &  4.9  &  33.4  &  3.9 & 29.1  &  2.6  &  25.2  &  \underline{0.9}  &  \textbf{19.8} &  1.0  &  \underline{20.0}\\
    Ice Ih (d)  &  \underline{0.8} & 25.4 &  3.2  &  29.2  &  2.6 & 24.1  &  3.2  &  43.4  &  \textbf{0.7}  &  \textbf{15.5} &  \underline{0.8}  &  \underline{17.9} \\\hline
    LWARMSE\tnote{a}  &  \textbf{0.8} & 33.0 &  3.4  &  44.8  &  2.5 & 38.0  &  1.6  &  33.9  &  \underline{1.0}  &  \textbf{25.2} &  \underline{1.0}  &  \underline{26.8} \\
    \toprule
  \end{tabular}
    \begin{tablenotes}
    \item[a] \footnotesize The Logarithmic Weighted Average RMSE defined in Eq.~\eqref{eq:LWARMSE}, with equal weights per subset.
  \end{tablenotes}
  \end{spacing}
  \end{threeparttable}
\end{table}

\begin{table}
  \caption{Test MAE on three datasets from Ref.~\cite{yin2025alphanet}. 
  Energy (E) MAE is in meV/atom (except in meV for Zeolite), force (F) MAE is in meV/\AA.} 
  \label{table:fcu}
  \centering  
  \begin{threeparttable}
  \begin{spacing}{1.3}
  \begin{tabular}{l | R{0.7cm}R{0.7cm} | R{0.7cm}R{0.7cm} | R{0.7cm}R{0.7cm} | R{0.7cm}R{0.7cm} | R{0.7cm}R{0.7cm}}
    \toprule
     & \multicolumn{2}{c|}{NequIP} & \multicolumn{2}{c|}{AlphaNet} & \multicolumn{2}{c|}{DPA3-L6} & \multicolumn{2}{c|}{DPA3-L12} & \multicolumn{2}{c}{DPA3-L24} \\\hline
    \textbf{Dataset} & \textbf{E~~~} & \textbf{F~~~} & \textbf{E~~~} & \textbf{F~~~} & \textbf{E~~~} & \textbf{F~~~} & \textbf{E~~~} & \textbf{F~~~} & \textbf{E~~~} & \textbf{F~~~} \\
    Formate D. on Cu  &  0.50 & 47.3  &  0.23  &  45.5  &  0.21 & 44.5 &  \underline{0.19} & \underline{39.7} &  \textbf{0.17} & \textbf{36.9}   \\
    Defected Graphene  &  1.90 & 60.2  &  1.70  &  32.0  &  0.72 & 35.2  &  \underline{0.62} & \underline{30.0} &  \textbf{0.60} & \textbf{28.2} \\
    Zeolite\tnote{a}  &  - & -  &  \textbf{15.4}  &  \textbf{14.6}  &  29.6 & 23.9  &  22.2 & 19.1 &  \underline{16.2} & \underline{15.1} \\
    \toprule
  \end{tabular}
  \begin{tablenotes}
    \item[a] \footnotesize MAE averaged via LWAMAE defined in Eq.~\eqref{eq:LWARMSE}, with equal weights per subset.
  \end{tablenotes}
  \end{spacing}
  \end{threeparttable}
\end{table}

\begin{table}
  \scriptsize
  \caption{Test RMSE on the DPA2 test sets which consist of 18 distinct datasets from Ref.~\cite{zhang2024dpa2}. 
  Energy (E) RMSE is in meV/atom and Force (F) RMSE is in meV/\AA across different models.} 
  \label{table:single_task}
  \centering  
  \begin{threeparttable}
  \begin{spacing}{1.3}
  \begin{tabular}{L{1.5cm} | c | R{0.44cm}R{0.44cm} | R{0.44cm}R{0.44cm} | R{0.44cm}R{0.44cm} | R{0.44cm}R{0.44cm} | R{0.44cm}R{0.44cm} | R{0.44cm}R{0.44cm} | R{0.44cm}R{0.44cm}}
    \toprule
    & \textbf{Weight} & \multicolumn{2}{c|}{GNO} & \multicolumn{2}{c|}{eqV2} & \multicolumn{2}{c|}{NequIP} & \multicolumn{2}{c|}{Allegro} & \multicolumn{2}{c|}{MACE} & \multicolumn{2}{c|}{DPA2} & \multicolumn{2}{c}{DPA3-L24} \\\hline
    \textbf{Dataset} & & \textbf{E~~~} & \textbf{F~~~} & \textbf{E~~~} & \textbf{F~~~} & \textbf{E~~~} & \textbf{F~~~} & \textbf{E~~~} & \textbf{F~~~} & \textbf{E~~~} & \textbf{F~~~} & \textbf{E~~~} & \textbf{F~~~} & \textbf{E~~~} & \textbf{F~~~} \\
    Alloy         & 2.0 & 14.3  & \underline{85.1}  & \underline{8.5}   & \textbf{62.7}  & 44.0  & 175.6 & 21.4  & 119.4 & 16.2  & 190.2 & 16.8  & 125.7 & \textbf{7.1}  & 99.2 \\
    Cathode-P     & 1.0  & 1.5   & \underline{17.9}  & 1.1   & \textbf{14.9}  & 14.3 & 69.8 & 1.0   & 24.2  & 2.6   & 37.8  & \underline{0.9}   & 24.5  & \textbf{0.6}  & 18.6  \\
    Cluster-P     & 1.0  & 47.7  & \textbf{69.6}  & 34.6  & \underline{104.4} & 75.1  & 216.6 & 54.8  & 174.1 & 41.3  & 189.7 & \underline{31.5}  & 126.0 & \textbf{29.3} & 118.3 \\
    Drug          & 2.0 & 40.5  & \underline{93.6}  & 29.8  & 807.4 & 21.6  & 187.2 & 13.1  & 100.8 & /~\tnote{b}     & /     & \underline{12.7}  & 125.5 & \textbf{5.5}  & \textbf{54.2}  \\
    FerroEle-P    & 1.0  & 1.5   & 17.9  & 1.1   & \textbf{13.0}  & 1.1   & 23.0  & 0.7   & 28.6  & 2.3   & 31.7  & \underline{0.6}   & 28.7  & \textbf{0.3}  & \underline{13.4}  \\
    OC2M         & 2.0   & 25.0  & 129.1 & \textbf{6.7}   & \textbf{45.2}  & 97.4  & 226.1 & 61.3  & 166.8 & /     & /     & 36.2  & 154.0 & \underline{9.0} & \underline{128.0} \\
    SSE-PBE-P      & 1.0 & 2.7   & \textbf{8.2}   & OOM   & OOM   & 1.6   & 41.1  & \underline{1.0}   & 47.8  & 1.8   & 29.9  & 1.4   & 50.3  & \textbf{0.5}  & \underline{19.8}  \\
    SemiCond      & 1.0  & 8.0   & \underline{94.4}  & \underline{3.9}   & \textbf{40.8}  & 20.5  & 180.7 & 6.8   & 146.8 & 12.7  & 182.8 & 5.5   & 123.6 & \textbf{3.4}  & 108.3 \\
    H2O-PD        & 1.0  & OOM\tnote{a}   & OOM   & OOM   & OOM   & 0.9   & 27.1  & OOM   & OOM   & 79.9  & 29.7  & \underline{0.5}   & \underline{24.7}  & \textbf{0.4}  & \textbf{13.7}  \\
    Ag$\cup$Au-PBE  & 0.2 & 106.0 & \underline{8.0}   & 23.4  & \textbf{4.4}   & 42.3  & 43.8  & 39.2  & 58.9  & 369.1 & 34.5  & \underline{2.4}   & 17.8  & \textbf{1.1}  & 10.9  \\
    Al$\cup$Mg$\cup$Cu  & 0.3 & 5.9 & \underline{9.4}   & \textbf{1.9}   & \textbf{5.7}   & 38.0  & 48.3  & 18.3  & 40.6  & 7.7   & 42.9  & 2.1   & 19.1  & \underline{2.0}  & 13.0  \\
    Cu          & 0.1    & 6.1   & \underline{5.8}   & 1.7   & \textbf{3.8}   & 6.2   & 16.7  & \underline{1.3}   & 8.9   & 38.8  & 13.6  & \textbf{1.2}   & 8.9   & 1.7  & 7.2   \\
    Sn            & 0.1  & 8.4   & \underline{33.7}  & 5.2   & \textbf{19.6}  & 18.2  & 62.2  & 5.6   & 40.2  & /     & /     & \underline{4.1}   & 54.4  & \textbf{2.9}  & 49.8  \\
    Ti           & 0.1   & 44.5  & 87.9  & 19.1  & \textbf{48.6}  & 27.6  & 137.4 & 6.9   & \underline{85.6}  & 8.3   & 94.2  & \underline{5.0}   & 113.1 & \textbf{4.1}  & 91.7  \\
    V            & 0.1   & 17.9  & 79.3  & 5.6   & \textbf{47.4}  & 8.8   & 91.6  & 4.2   & 82.1  & 14.2  & 140.4 & \underline{4.1}   & 90.8  & \textbf{2.8}  & \underline{71.8}  \\
    W            & 0.1   & 79.1  & \underline{81.2}  & 46.8  & \textbf{51.3}  & 20.8  & 160.4 & \underline{4.0}   & 101.6 & 15.6  & 181.2 & 5.6   & 108.1 & \textbf{2.5}  & 83.3  \\
    C12H26        & 0.1  & 135.8 & \textbf{518.7} & 123.1 & 907.4 & 121.4 & 715.6 & 140.4 & 648.1 & 81.9  & 802.3 & \underline{55.3}  & 692.5 & \textbf{42.4} & \underline{541.6} \\
    HfO2        & 0.1    & \underline{1.2}   & 16.1  & \textbf{1.0}   & \textbf{9.1}   & 1.5   & 58.8  & 1.4   & 64.0  & 2.3   & \underline{14.7}  & \textbf{1.0}   & 54.2  & 1.4  & 28.9  \\\hline
    \multicolumn{2}{c|}{LWARMSE\tnote{c}} & 12.1 & \underline{52.0} & 7.4 & 62.0 & 9.3 & 87.2 & 16.5 & 107.2 & 11.3 & 75.3 & \underline{5.9} & 76.6 & \textbf{2.9} & \textbf{49.8} \\
    \toprule
  \end{tabular}
  \begin{tablenotes}
    \item[a] \footnotesize OOM indicates Out-Of-Memory errors.
    \item[b] \footnotesize / signifies an unresolved error that occurred during the training process.
    \item[c] \footnotesize The Logarithmic Weighted Average RMSE defined in Eq.~\eqref{eq:LWARMSE} across all systems, with weights defined in Ref.~\cite{zhang2024dpa2}. 
  \end{tablenotes}
  \end{spacing}
  \end{threeparttable}
\end{table}

\begin{table}
\scriptsize
\renewcommand{\arraystretch}{0.7}
\caption{Zero-shot generalizability for force-field prediction across three domains. 
Energy (E) RMSE is reported in meV/atom, Force (F) RMSE in meV/\AA, and Virial (V) RMSE in meV/atom across different models.
\recheck{Training data summary:
DPA-3.1-3M and DPA-2.4-7M~\cite{zhang2024dpa2,peng2025lambench} are multitask-pretrained on OpenLAM-v1~\cite{openlam-data-v1-web,peng2025lambench}.
SevenNet-MF-ompa~\cite{kim2024data} is multitask-pretrained on OMat24~\cite{barroso2024open}, MPtrj~\cite{deng2023chgnet}, and sAlex~\cite{schmidt2023machine}.
MACE-MPA-0~\cite{batatia2023foundation}, Orb-v3~\cite{rhodes2025orbv3}, and GRACE-2L-OAM~\cite{bochkarev2024graph} are pretrained on OMat24 and subsequently fine-tuned on MPtrj and sAlex.
MatterSim-v1-5M~\cite{yang2024mattersim} is trained on a proprietary materials dataset that is not publicly available.
}}
\centering
  \begin{threeparttable}
  \begin{spacing}{1.2}
\begin{tabular}{lllrrrrrrrr}
\toprule
\textbf{Domain} & \textbf{Dataset} & & \rotatebox{90}{\textbf{Dummy\tnote{a}}} & \rotatebox{90}{\textbf{DPA-3.1-3M (OMat)}} & \rotatebox{90}{\textbf{DPA-2.4-7M}} & \rotatebox{90}{\textbf{MACE-MPA-0}} & \rotatebox{90}{\textbf{Orb-v3}} & \rotatebox{90}{\shortstack[c]{\textbf{SevenNet-MF-ompa}\\\textbf{(OMat)}}}
 & \rotatebox{90}{\textbf{MatterSim-v1-5M}} & \rotatebox{90}{\textbf{GRACE-2L-OAM}} \\
\midrule
Catalysis & Vandermause2022Active~\cite{Vandermause2022} & \textbf{E~~~} & 48.9 & 5.7 & 14.2 & 10.4 & 3.0 & 13.4 & 13.5 & 5.5 \\ &  & \textbf{F~~~} & 1164.1 & 85.1 & 151.6 & 175.0 & 89.4 & 99.4 & 157.7 & 99.4 \\ &  & \textbf{V~~~} & 270.7 & 34.2 & 104.3 & 71.1 & 34.7 & 27.9 & 75.2 & 60.6 \\ \cmidrule(l){2-11} & Zhang2019Bridging~\cite{doi:10.1021/acs.jpclett.9b00085} & \textbf{E~~~} & 11.7 & 4.2 & 10.7 & 93.8 & 288.5 & 367.9 & 562.8 & 251.6 \\ &  & \textbf{F~~~} & 691.5 & 56.9 & 130.6 & 351.6 & 968.7 & 897.6 & 1651.4 & 723.0 \\ \cmidrule(l){2-11} & Villanueva2024Water~\cite{doi:10.1021/jacs.3c10685} 
& \textbf{E~~~} & 14.2 & 3.4 & 4.1 & 4.4 & 2.2 & 2.9 & 3.7 & 3.3 \\ 
&  & \textbf{F~~~} & 187.1 & 120.1 & 121.8 & 142.6 & 81.1 & 99.3 & 112.6 & 131.8 \\
\cmidrule(l){2-11} 
& \textbf{LWARMSE\tnote{b}} 
& \textbf{E~~~} & / & \textbf{4.3} & \underline{8.5} & 16.2 & 12.4 & 24.2 & 30.3 & 16.6 \\ 
&  & \textbf{F~~~} & / & \textbf{83.4} & \underline{134.1} & 206.3 & 191.5 & 206.9 & 308.4 & 211.5 \\ 
&  & \textbf{V~~~} & / & \underline{34.2} & 104.3 & 71.1 & 34.7 & \textbf{27.9} & 75.2 & 60.6 \\
\midrule Inorganic Materials & Torres2019Analysis~\cite{Ca_batteries_CM2021} & \textbf{E~~~} & 15.7 & 2.7 & 3.8 & 2.7 & 2.8 & 2.8 & 2.5 & 3.2 \\ &  & \textbf{F~~~} & 176.3 & 166.1 & 152.8 & 181.9 & 153.5 & 167.1 & 169.8 & 161.2 \\ \cmidrule(l){2-11} & Batzner2022equivariant~\cite{batzner20223} & \textbf{E~~~} & 7.0 & 0.8 & 1.6 & 1.0 & 0.6 & 0.7 & 1.5 & 0.8 \\ &  & \textbf{F~~~} & 943.5 & 86.5 & 138.7 & 117.6 & 78.2 & 79.1 & 106.0 & 111.5 \\ \cmidrule(l){2-11} & Gao2025Spontaneous~\cite{Gao2025} & \textbf{E~~~} & 81.7 & 17.5 & 22.5 & 29.0 & 14.4 & 14.9 & 15.2 & 17.8 \\ &  & \textbf{F~~~} & 956.8 & 103.3 & 182.0 & 193.5 & 63.9 & 104.1 & 122.2 & 135.7 \\ &  & \textbf{V~~~} & 416.2 & 106.6 & 225.0 & 322.0 & 76.1 & 78.5 & 145.1 & 164.1 \\ \cmidrule(l){2-11} & Sours2023Applications~\cite{doi:10.1021/acs.jpcc.2c08429} 
& \textbf{E~~~} & 35.3 & 6.6 & 7.0 & 6.5 & 6.6 & 6.6 & 6.7 & 6.5 \\ 
&  & \textbf{F~~~} & 1232.5 & 174.0 & 186.5 & 170.6 & 176.2 & 175.8 & 182.6 & 162.8 \\
\cmidrule(l){2-11} & Lopanitsyna2023Modeling~\cite{PhysRevMaterials.7.045802} & \textbf{E~~~} & 510.8 & 51.9 & 59.2 & 83.6 & 61.2 & 55.8 & 72.9 & 70.1 \\ &  & \textbf{F~~~} & 2262.6 & 203.9 & 290.3 & 363.4 & 272.6 & 243.7 & 343.6 & 309.7 \\ &  & \textbf{V~~~} & 2350.6 & 164.2 & 196.5 & 292.8 & 265.9 & 186.2 & 233.6 & 243.1 \\ \cmidrule(l){2-11} & Mazitov2024Surface~\cite{Mazitov_2024} & \textbf{E~~~} & 431.2 & 42.8 & 69.3 & 61.3 & 58.1 & 49.5 & 75.4 & 55.4 \\ &  & \textbf{F~~~} & 1645.2 & 211.3 & 274.2 & 350.4 & 285.7 & 247.4 & 308.4 & 294.7 \\ &  & \textbf{V~~~} & 1534.7 & 154.7 & 180.5 & 257.1 & 204.7 & 178.9 & 189.7 & 200.1 \\ \cmidrule(l){2-11} & 
\textbf{LWARMSE} 
& \textbf{E~~~} & / & \underline{9.1} & 12.5 & 11.7 & \textbf{9.0} & \textbf{9.0} & 11.3 & 10.2 \\ 
&  & \textbf{F~~~} & / & \underline{149.4} & 196.3 & 211.6 & \textbf{148.0} & 156.3 & 186.9 & 181.9 \\ 
&  & \textbf{V~~~} & / & \underline{139.4} & 199.8 & 289.4 & 160.6 & \textbf{137.8} & 185.9 & 199.9 \\ 
\midrule Molecules & MD22~\cite{doi:10.1126/sciadv.adf0873} 
& \textbf{E~~~} & 7.9 & 2.6 & 2.9 & 7.1 & 3.4 & 4.3 & 3.7 & 3.5 \\ 
&  & \textbf{F~~~} & 1139.1 & 129.6 & 149.2 & 247.2 & 170.1 & 239.3 & 257.2 & 235.9 \\
\cmidrule(l){2-11} & AIMD-Chig~\cite{Wang2023} & \textbf{E~~~} & 9.4 & 2.6 & 2.7 & 5.1 & 2.3 & 3.5 & 3.3 & 3.1 \\ &  & \textbf{F~~~} & 870.0 & 152.8 & 170.9 & 289.4 & 196.9 & 259.2 & 281.2 & 239.8 \\ \cmidrule(l){2-11} 
& ANI-1x~\cite{10.1063/1.5023802} & \textbf{E~~~} & 191.5 & 20.4 & 23.6 & 37.2 & 23.2 & 26.4 & 30.5 & 32.1 \\ &  & \textbf{F~~~} & 2078.5 & 191.4 & 218.9 & 382.0 & 253.0 & 328.5 & 341.5 & 365.5 \\ \cmidrule(l){2-11} 
& \textbf{LWARMSE} 
& \textbf{E~~~} & / & \textbf{5.2} & \underline{5.6} & 11.1 & \underline{5.6} & 7.4 & 7.2 & 7.0 \\ 
&  & \textbf{F~~~} & / & \textbf{155.9} & \underline{177.4} & 301.2 & 203.9 & 273.1 & 291.2 & 274.5 \\ 
\midrule 
& & \textbf{E~~~} & /  & \textbf{6.6} & 9.3 & 12.5 & \underline{8.7} & 11.0 & 12.9 & 10.5 \\ \textbf{LWARMSE} & & \textbf{F~~~} & /  & \textbf{130.6} & 174.0 & 229.7 & \underline{171.0} & 192.8 & 236.6 & 209.4 \\ & & \textbf{V~~~} & /  &  \underline{98.1} &  169.9 &  203.7 &  109.4 &  \textbf{92.4} &  148.3 &  148.3 \\
\bottomrule
\end{tabular}
  \begin{tablenotes}
    \item[a] \footnotesize The dummy linear model that predicts energies from chemical composition and outputs zero forces and virials, used to estimate the standard deviation for each dataset.
    \item[b] \footnotesize The Logarithmic Weighted Average RMSE defined in Eq.~\eqref{eq:LWARMSE}, with equal weights per dataset.
  \end{tablenotes}
  \end{spacing}
  \end{threeparttable}
\label{tab:ffperformance}
\end{table}

\begin{table}
  \scriptsize
  \caption{\recheck{Non-compliant model results on the Matbench Discovery leaderboard, with all models discussed in Sec.~\ref{sec:LAM} and accessed before
  May 27, 2025. DPA-3.1-3M-FT denotes DPA-3.1-3M finetuned on the MPtrj and sAlex datasets, with training stopped when the energy error no longer decreased appreciably.}}
  \label{table:sm:noncom:matbench}
  \centering  
  \begin{threeparttable}
  \begin{spacing}{1.3}
  \begin{tabular}{L{2.4cm}R{0.7cm}R{0.7cm}R{0.7cm}R{0.7cm}R{0.7cm}R{0.7cm}R{0.7cm}R{0.7cm}R{0.8cm}R{0.8cm}R{0.8cm}R{0.8cm}}
    \toprule
    \textbf{Model} & \textbf{CPS}$\uparrow$ & \textbf{Acc}$\uparrow$ & \textbf{F1}$\uparrow$ & \textbf{DAF}$\uparrow$ & \textbf{Prec}$\uparrow$ & \textbf{MAE}$\downarrow$ & 
    \textbf{RMSE}$\downarrow$ & \textbf{R2}$\uparrow$ & \textbf{$\kappa$SRME}$\downarrow$ & \textbf{RMSD}$\downarrow$ & \textbf{Params}$\downarrow$ & \textbf{Targets} \\\hline
    DPA-3.1-3M-FT & 0.802 & 0.963 & 0.884 & 5.667 & 0.866 & 0.023 & \textbf{0.067} & \textbf{0.869} & 0.469 & 0.069 & \textbf{3.27M} & EFSG \\
    \hline
    eSEN-30M-OAM & \textbf{0.888} & \textbf{0.977} & \textbf{0.925} & \textbf{6.069} & \textbf{0.928} & \textbf{0.018} & \textbf{0.067} & 0.866 & \textbf{0.170} & \textbf{0.061} & 30.2M & EFSG \\
    Orb-v3 & \underline{0.861} & \underline{0.971} & \underline{0.905} & \underline{5.912} & \underline{0.904} & 0.024 & 0.078 & 0.821 & \underline{0.210} & 0.075 & 25.5M & EFSG \\
    SevenNet-MF-ompa & 0.845 & 0.969 & 0.901 & 5.825 & 0.89 & \underline{0.021} & \textbf{0.067} & \underline{0.867} & 0.317 & \underline{0.064} & 25.7M & EFSG \\
    GRACE-2L-OAM & 0.837 & 0.963 & 0.880 & 5.774 & 0.883 & 0.023 & 0.068 & 0.862 & 0.294 & 0.067 & 12.6M & EFSG \\
    MACE-MPA-0 & 0.795 & 0.954 & 0.852 & 5.582 & 0.853 & 0.028 & 0.073 & 0.842 & 0.412 & 0.073 & 9.06M & EFSG \\
    MatterSim-v1-5M & 0.767 & 0.959 & 0.862 & 5.852 & 0.895 & 0.024 & 0.068 & 0.863 & 0.574 & 0.073 & \underline{4.55M} & EFSG \\
    \toprule
  \end{tabular}
  \end{spacing}
  \end{threeparttable}
\end{table}

\begin{table}
  \caption{\recheck{Test MAE on SPICE-MACE-OFF dataset. DPA-3.1-3M is finetuned with the SPICE2 dataset encoding, using the same settings as the second training round of the task-specific variants. Results for other models are provided in Table~\ref{table:spice}. Energy (E) MAE is in meV/atom and force (F) MAE is in meV/\AA.}}
  \label{table:sm:spice:ft}
  \centering
  \begin{threeparttable}
  \begin{spacing}{1.15}
  \begin{tabular}{l
    | R{1cm}R{1cm}  
    | R{1cm}R{1cm}  
    | R{1cm}R{1cm}  
  }
    \toprule
     & \multicolumn{2}{c|}{DPA3-L12}
     & \multicolumn{2}{c|}{DPA3-L24}
     & \multicolumn{2}{c}{\begin{tabular}{@{}c@{}}Fine-tuned\\DPA-3.1-3M\end{tabular}} \\\hline
    \textbf{Dataset} & \textbf{E~~~} & \textbf{F~~~} & \textbf{E~~~} & \textbf{F~~~} & \textbf{E~~~} & \textbf{F~~~} \\
    PubChem      & 0.41 & 12.12 & 0.24 &  8.47 & 0.42 & 13.66 \\
    DES370K M.   & 0.31 &  5.25 & 0.18 &  3.15 & 0.26 &  5.63 \\
    DES370K D.   & 0.43 &  5.13 & 0.23 &  3.19 & 0.38 &  5.45 \\
    Dipeptides   & 0.24 &  7.55 & 0.13 &  4.81 & 0.22 &  8.28 \\
    Sol. AA      & 0.38 & 12.69 & 0.31 &  8.77 & 0.39 & 13.95 \\
    Water        & 0.47 &  9.83 & 0.32 &  6.89 & 0.60 & 10.90 \\
    QMugs        & 0.32 & 12.90 & 0.17 &  8.66 & 0.29 & 14.27 \\\hline
    LWAMAE\tnote{a} & 0.36 &  8.74 & 0.22 &  5.78 & 0.35 &  9.60 \\\hline
    \textbf{\# Params} & \multicolumn{2}{c|}{2.5 M} & \multicolumn{2}{c|}{4.9 M} & \multicolumn{2}{c}{3.3 M} \\
    \toprule
  \end{tabular}

  \begin{tablenotes}
    \item[a] \footnotesize The Logarithmic Weighted Average MAE defined in Eq.~\eqref{eq:LWARMSE}.
  \end{tablenotes}
  \end{spacing}
  \end{threeparttable}
\end{table}




\section{Model training}
\subsection{Robustness of activation function}
\label{sec:activation}
During the training of the DPA3 model, we found that certain normalization techniques (such as layer normalization or batch normalization) had a negative impact on both accuracy and efficiency. 
Consequently, explicit normalization was not incorporated into the model architecture. 

On the other hand, we employed the $\mathrm{SiLU}$ activation function, which outperformed $\mathrm{tanh}$ in terms of accuracy; however, it is an unbounded activation function. 
We observed that during training, particularly with deeper networks, there is a tendency for numerical instability leading to explosive accumulation of values. 
Therefore, we replaced $\mathrm{SiLU}$ with our new activation function, termed $\mathrm{SiLUT}$ ($\mathrm{SiLU}$ threshold with $\mathrm{Tanh}$), defined as follows:
\begin{align}
    \mathrm{SiLUT}(x) = 
    \begin{cases}
    \mathrm{SiLU}(x) & \text{if } x \leq t,  \\ 
    \mathrm{tanh}(a \cdot(x-t)) + b & \text{if } x > t,
\end{cases}
\end{align}
where $t$ represents the threshold, indicating when to transition from SiLU to tanh.
Constants $a$ and $b$ are determined based on $t$ to ensure the first and second continuity of the activation function at the threshold.

We found that this activation function design, in most cases, maintains the accuracy of SiLU while providing more stable training, especially in the absence of explicit normalization.

\subsection{Model hyperparameters}
\label{sec:hyperparameters}
The hyperparameters utilized for training the DPA3 models are summarized in Table~\ref{table:hyperparam}. 
All models were trained using a consistent architecture and fixed dimensions for vertex and edge features across different graph orders. 
To ensure optimal utilization and balancing across different devices, especially using multiple GPUs, batch sizes were dynamically determined based on the number of atoms present in each specific system. 
An exponential decay strategy was employed for learning rate scheduling throughout the training process, with simultaneous adjustments to the prefactors of loss components in synchronization with these learning rate changes. 
For both the SPICE-MACE-OFF and MPtrj datasets, training was conducted in two rounds, where the second round was initialized from the checkpoint of the first. 
Each round employed distinct configurations of loss functions and prefactors, leading to enhanced overall performance.
\recheck{
For DPA-3.1-3M, we performed multitask pretraining on OpenLAM-v1 using the same architecture as the task-specific DPA3 variants. 
The corresponding hyperparameters are also listed in Table~\ref{table:hyperparam}.
}

\begin{table}
\scriptsize
\caption{Hyper-parameters for DPA3 models trained on different benchmark datasets.}
\label{table:hyperparam}
\centering
\begin{threeparttable}
\begin{spacing}{1.3}
 \begin{tabular}{l|cccccccc}
  \toprule
   \textbf{Hyper-parameters} & \begin{tabular}[c]{@{}c@{}}SPICE\\-round1\end{tabular} & \begin{tabular}[c]{@{}c@{}}SPICE\\-round2\end{tabular} & \begin{tabular}[c]{@{}c@{}}MPtrj\\-round1\end{tabular} & \begin{tabular}[c]{@{}c@{}}MPtrj\\-round2\end{tabular} & Scaling-law & Graph-order & DPA-3.1-3M & Others \\\hline 
   Number of update layers $L$ & \recheck{3/}6/12/24 & \recheck{3/}6/12/24 & 24 & 24 & 3/6/9/16/24 & 6/12/24 & 16 & 6/12/24 \\
   Maximum graph order $k$ & 2 & 2 & 2 & 2 & 2 & 3 & 2 & 2 \\
   Dimension of vertex features in $G^{(1)}$ & 128 & 128 & 128 & 128 & 128 & 128 & 128 & 128 \\
   Dimension of edge features in $G^{(1)}$ & 64 & 64 & 64 & 64 & 64 & 64 & 64 & 64 \\
   Dimension of edge features in $G^{(2)}$ & 32 & 32 & 32 & 32 & 32 & 32 & 32 & 32 \\
   Dimension of edge features in $G^{(3)}$ & - & - & - & - & - & 32 & - & - \\
   Cutoff radius in $G^{(1)}$ (\AA) & 6.0 & 6.0 & 6.0 & 6.0 & 6.0 & 6.0 & 6.0 & 6.0 \\
   Cutoff radius in $G^{(2)}$ (\AA) & 4.0 & 4.0 & 4.5 & 4.5 & 4.0 & 4.0 & 4.0 & 4.0 \\
   Cutoff radius in $G^{(3)}$ (\AA) & - & - & - & - & - & 2.8 & - & - \\
   Batch size (per GPU) &  $\lceil \frac{256}{N} \rceil$\tnote{a} & $\lceil \frac{256}{N} \rceil$ & $\lceil \frac{128}{N} \rceil$ & $\lceil \frac{128}{N} \rceil$ & $\lceil \frac{128}{N} \rceil$ & 1 & $\lceil \frac{128}{N} \rceil$ & 1 or $\lceil \frac{256}{N} \rceil$\tnote{e} \\
   Number of GPUs & 8 & 8 & 16 & 16 & 16 & 1 & 128 & 1 \\
   Optimizer & Adam & Adam & AdamW & AdamW & AdamW & Adam & Adam & Adam \\
   Learning rate scheduling & Exp & Exp & Exp & Exp & Exp & Exp & Exp & Exp \\
   Maximum learning rate & 1e-3 & 1e-3 & 1e-3 & 5e-4 & 1e-3 & 1e-3 & 1e-3 & 1e-3 \\
   Minimum learning rate  & 1e-5 & 1e-5 & 1e-5 & 1e-5 & 1e-5 & 1e-5 & 1e-6 & 1e-5 \\
   Number of training steps & 4M & 4M & 2M\tnote{c} & 2M\tnote{c} & \recheck{-\tnote{d}} & 1M & 4M & 1M \\
   Activation Function Threshold & 10.0 & 5.0 & 10.0 & 5.0 & 10.0 & 10.0 & 3.0 & 10.0 \\
   Loss function  & MSE & Huber & MSE & Huber & MSE & MSE & MSE & MSE \\
   Energy loss prefactor & 0.2$\rightarrow$\tnote{b}~~20 & 15 & 0.2$\rightarrow$20 & 15 & 0.2$\rightarrow$20 & 0.2$\rightarrow$20 & 0.2$\rightarrow$20 & 0.2$\rightarrow$20 \\
   Force loss prefactor & 100$\rightarrow$60 & 1 & 100$\rightarrow$20 & 1 & 100$\rightarrow$20 & 100$\rightarrow$20 & 100$\rightarrow$60 & 100$\rightarrow$20 \\
   Virial loss prefactor & - & - & 0.02$\rightarrow$1 & 2.5 & 0.02$\rightarrow$1 & 0.02$\rightarrow$1 & 0.02$\rightarrow$1 & 0.02$\rightarrow$1 \\
   \toprule
 \end{tabular}
 \begin{tablenotes}
    \item[a] \footnotesize $N$ denotes the number of atoms in each system, $\lceil \cdot \rceil$ denotes the ceiling function, which rounds the number up to the nearest integer.
    \item[b] \footnotesize $\rightarrow$ indicates that the prefactors are changed in synchronization with the learning rate.
    \item[c] \footnotesize We use early stopping for MPtrj training based on the validation energy MAE on the WBM test set.
    \item[d] \footnotesize \recheck{Depends on compute resource settings}.
    \item[e] \footnotesize We use batchsize=1 for the DPA2 test sets for fair comparison and $\lceil \frac{256}{N} \rceil$ for other benchmark tests.
 \end{tablenotes}
 \end{spacing}
 \end{threeparttable}
\end{table}

\section{Ablation study}
\subsection{Ablation for different LiGS orders and cutoffs}
\recheck{
To assess how LiGS orders and per-order cutoffs affect DPA3, we trained DPA3–L24 variants on MPtrj dataset (8 GPUs, 1M steps) and report energy MAE on the WBM dataset. 
For efficiency, we averaged inference time over 100 frames randomly sampled from MPtrj. 
As shown in Table.~\ref{table:sm:rcut}, with $G^{(1)}$ only, increasing the first-order cutoff ($r_{c}^1$) from 4 to 6 reduces MAE by 30\%, indicating a strong accuracy impact. 
Building on $G^{(1)}$ with $r_{c}^1$ = 6, adding a second-order graph $G^{(2)}$ with $r_{c}^2$ = 4 provides a further 12\% MAE reduction, but inference speed decreases by roughly 3x due to higher computational cost. 
Note that the absolute efficiency here is higher than in Fig.~\ref{fig:sm:effi} because MPtrj contains smaller systems on average (around 30 atoms).
Modestly enlarging $r_{c}^2$ to 4.5 improves accuracy again, at the expense of additional slowdowns. 
These observations motivate using smaller cutoffs at higher orders, reflecting a practical accuracy–efficiency trade-off.
}

\begin{table}
  \caption{\recheck{Effect of truncated LiGS orders and per–order cutoffs on DPA3.
  DPA3–L24 variants are trained on MPtrj with 8 GPU cards for 1M steps and evaluated on WBM. 
  Inference efficiency is averaged over 100 randomly selected MPtrj frames.
  $r_{c}^1$ and $r_{c}^2$ denote the cutoff radius in $G^{(1)}$ and $G^{(2)}$, respectively.
  }
  } 
  \label{table:sm:rcut}
  \centering  
  \begin{threeparttable}
  \begin{spacing}{1.3}
  \begin{tabular}{l | R{2cm} | R{1.5cm} | R{1.5cm} | R{2.5cm} | R{2.5cm}}
    \toprule
    Model & LiGS order & $r_{c}^1$ ~(\AA) & $r_{c}^2$ ~(\AA) & Energy MAE (meV/atom) & Efficiency (ms/atom)\\\hline
    \textbf{DPA3-L24}  & 1 & 4.0 & /   & 56.1 & 0.3\\
                       & 1 & 6.0 & /   & 39.5 & 0.4\\
                       & 2 & 6.0 & 4.0 & 34.7 & 1.3\\
                       & 2 & 6.0 & 4.5 & 32.2 & 1.8\\
    \toprule
  \end{tabular}
  \end{spacing}
  \end{threeparttable}
\end{table}

\subsection{Ablation for model architecture}
\label{sec:sm:ablation}
\begin{figure}
    \centering
    \includegraphics[width=0.5\linewidth]{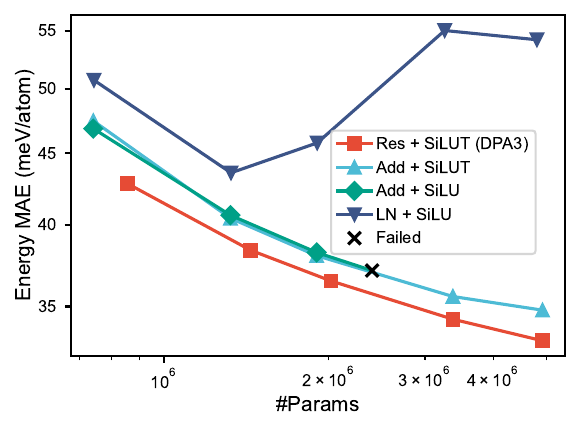}
    \caption{
    \recheck{Ablation of DPA3 substructures on the MPtrj dataset across different parameter counts.
All models were trained for 1M steps using eight GPUs.
Energy MAE (meV/atom) is reported on the WBM test set.
Res denotes the residual update (Eqs.~\eqref{eq:nupdate} and \eqref{eq:eupdate}); Add denotes the additive update (Eqs.~\eqref{eq:addnupdate} and \eqref{eq:addeupdate}); SiLUT refers to the custom activation function (Sec.~\ref{sec:activation}); LN denotes the layer-normalized update (Eqs.~\eqref{eq:lnnupdate} and \eqref{eq:lneupdate}).
“Failed” indicates training runs terminated due to numerical instabilities.
    }
    }
    \label{fig:sm:ablation}
\end{figure}

\recheck{To investigate the influence of architectural substructures on DPA3 performance, we conducted ablation studies.
Models with varying architectural choices and numbers of layers were trained on the MPtrj dataset for a fixed 1M training steps using eight GPUs.
The energy MAEs on the WBM test set were evaluated and are reported in Fig.~\ref{fig:sm:ablation}.

We compared three types of feature updates: the residual update used by DPA3 (Eqs.~\eqref{eq:nupdate}–\eqref{eq:eupdate}, denoted as “Res”), the direct additive update adopted by ALIGNN-FF~\cite{choudhary2021atomistic} (denoted as “Add”),
\begin{align}
\label{eq:addnupdate}
& v_\alpha^{(k,l+1)} \gets v_\alpha^{(k,l+1)} + u_\alpha^{(k,l)}, \\
\label{eq:addeupdate}
& e_{\alpha\beta}^{(k,l+1)} \gets e_{\alpha\beta}^{(k,l+1)} + m_{s,\alpha\beta}^{(k,l)},
\end{align}
and the layer-normalized update used by CHGNet~\cite{deng2023chgnet}  (denoted as “LN”):
\begin{align}\label{eq:lnnupdate}
& v_\alpha^{(k,l+1)} \gets \operatorname{LayerNorm}\big(v_\alpha^{(k,l+1)} + u_\alpha^{(k,l)}\big), \\\label{eq:lneupdate}
& e_{\alpha\beta}^{(k,l+1)} \gets \operatorname{LayerNorm}\big(e_{\alpha\beta}^{(k,l+1)} + m_{s,\alpha\beta}^{(k,l)}\big).
\end{align}
We also evaluated the SiLUT activation function relative to the SiLU activation used in ALIGNN-FF and CHGNet.

Four ablation experiments were conducted: the baseline Res update + SiLUT (DPA3), Add update + SiLUT, Add update + SiLU, and LN update + SiLU.
Replacing the Res update with Add preserved the error-scaling behavior (rate of MAE reduction with model size) while slightly reducing accuracy.
Using SiLU instead of SiLUT on the Add update maintained accuracy for small models but introduced numerical instabilities for models exceeding $2\times10^6$ parameters (corresponding to nine layers), indicating that the SiLUT activation enhances stability for deeper models.
Finally, applying layer normalization stabilized training, but error scaling stopped at approximately 1.3M parameters (six layers), and further increasing depth led to higher test errors.

}

\subsection{Dataset encoding in multitask training}
\label{sec:sm:encoding}
\recheck{To evaluate the effect of dataset encoding versus separate fitting heads, we performed multitask training following the DPA-3.1-3M setup, but with reduced training steps and compute budgets, as comparable compute resources to the full DPA-3.1-3M training were not available.
All other hyperparameters were kept consistent with DPA-3.1-3M.
As shown in Table~\ref{tab:sepshare}, dataset encoding achieves nearly identical energy, force, and virial accuracies to separate fitting networks while significantly reducing the parameter count.

}

\begin{table}
\scriptsize
\caption{Accuracy comparison between multi-task training with dataset encoding and separate fitting networks under reduced training steps and GPU budgets.
The overall Logarithmic Weighted Average RMSE (LWARMSE) for Energy (E, meV/atom), Force (F, meV/\AA), and Virial (V, meV/atom) from the zero-shot generalizability evaluation in Table~\ref{tab:ffperformance} are reported.
}
\centering
  \begin{threeparttable}
  \begin{spacing}{1.2}
\begin{tabular}{crrr|ccc}
\toprule
\textbf{Multi-task Mode} & \textbf{E~~~} & \textbf{F~~~} & \textbf{V~~~} & \# Training Steps & \# GPUs & \# Params\\
\midrule
Dataset encoding &  6.6 &  130.6  & 98.1 & 4M & 128 & 3.26M\\
\midrule
Dataset encoding &   11.4 &  205.8  & 131.9 & 1M & 8 & 3.26M\\
Seperate fitting networks &  11.6 &  211.0  &  127.9 & 1M & 8 & 7.69M\\ 
\bottomrule
\end{tabular}
  \end{spacing}
  \end{threeparttable}
\label{tab:sepshare}
\end{table}

\bibliography{ref}



\end{document}